\documentclass[amsmath,amssymb,superscriptaddress,nobalancelastpage,aps,prl,twocolumn]{revtex4-2}
\usepackage{graphicx}
\usepackage{varioref}
\usepackage{xr-hyper}
\usepackage{xcolor}
\usepackage{nicefrac}
\usepackage{xfrac}
\usepackage{hyperref}
\hypersetup{colorlinks,linkcolor=blue,urlcolor=blue,citecolor=blue}
\usepackage{ulem}
\usepackage{lineno}
\usepackage{amsmath} 
\usepackage{amssymb}
\usepackage{siunitx}

\begin{document}
	
\title{Spin correlations in the parent phase of \texorpdfstring{Li$_{1-x}$Fe$_x$ODFeSe}{}}
\author{Hongliang Wo$^\dagger$}
\affiliation{State Key Laboratory of Surface Physics and Department of Physics, Fudan University, Shanghai 200433, China}
\author{Bingying Pan$^\dagger$}
\affiliation{State Key Laboratory of Surface Physics and Department of Physics, Fudan University, Shanghai 200433, China}
\affiliation{College of Physics and Optoelectronic Engineering, Ocean University of China, Qingdao, Shandong 266100, China}
\author{Die Hu}
\affiliation{State Key Laboratory of Surface Physics and Department of Physics, Fudan University, Shanghai 200433, China}
\author{Yu Feng}
\affiliation{State Key Laboratory of Surface Physics and Department of Physics, Fudan University, Shanghai 200433, China}
\affiliation{Institute of High Energy Physics, Chinese Academy of Sciences (CAS), Beijing 100049, China}
\affiliation{Spallation Neutron Source Science Center, Dongguan 523803, China}
\author{A. D. Christianson}
\affiliation{Materials Science and Technology Division, Oak Ridge National Laboratory, Oak Ridge, Tennessee 37831, USA}
\author{Jun Zhao$^\ast$}
\affiliation{State Key Laboratory of Surface Physics and Department of Physics, Fudan University, Shanghai 200433, China}
\affiliation{Shanghai Research Center for Quantum Sciences, Shanghai 201315, China}
\affiliation{Institute of Nanoelectronics and Quantum Computing, Fudan University, Shanghai 200433, China}

\begin{abstract}
{Elucidating spin correlations in the parent compounds of high-temperature superconductors is crucial for understanding superconductivity. We used neutron scattering to study spin correlations in Li$_{1-x}$Fe$_x$ODFeSe, an insulating material with reduced electron carriers compared to its superconducting counterpart (\textit{T}$_c$ = 41 K), serving as the undoped parent compound. Our findings show a reduced total fluctuating moment in this insulator relative to FeSe and 122 iron pnictides, likely due to increased interlayer distances from intercalation,  which enhance fluctuations and reduce the intensity of spin excitations. Moreover, we observed a V-shaped spin wave-like excitation dispersion, contrasting with the twisted hourglass pattern in the superconducting counterpart. Electron doping shifts spin excitation from ($\pi$, 0) point to an incommensurate position towards ($\pi$, $\pi$) direction below 65 meV. This transition from V-shaped to hourglass-like dispersion, akin to behaviors in hole-doped cuprates, suggests a potential shared mechanism in magnetism and superconductivity across these diverse systems. }
\end{abstract}
\maketitle

Iron-based superconductivity emerges when sufficient carriers are introduced to an antiferromagnetic semi-metallic parent compound through chemical doping \cite{Hosono}. Understanding the inherent spin fluctuations in the parent compounds, as well as how they evolve with carrier doping, is essential to unravel the mysterious pairing mechanism, as spin fluctuations may play a key role in mediating Cooper pairing \cite{DaiRev,ScalapinoRev}. Earlier measurements have revealed that carrier doping in iron pnictides leads to a broadening of the parent compound's spin wave excitation spectrum and induces a resonance mode below the superconducting gap \cite{DaiRev}. These spin excitations are believed to be the driving forces behind an \textit{s}-wave pairing, characterized by a sign-reversal between electron and hole pockets in the superconducting gap function \cite{DaiRev,ScalapinoRev}.

Electron-doped iron selenide superconductors have recently attracted substantial attention due to their unique electronic structure and notably high critical temperature (\textit{T}$_c$) \cite{KFe2Se2,FeSe_film_Tc1,XianhuiNM,LiFeO_41K}. The Fermi surface topology of these superconductors diverges from that found in iron pnictides, containing only electron pockets at the corners of the Brillouin zone, and devoid of hole pockets at the zone center \cite{FS_AFeSe1,FS_AFeSe2,FS_AFeSe3,FS_AFeSe4,FS_FeSe1,FS_FeSe2,QMSi_Review,FS_LiFeO1,FS_LiFeO2,FS_LiFeO3,FS_LiFeO4}. This is incompatible with the proposed sign-reversed \textit{s}-wave Cooper pairing mechanism associated with the scattering between electron and hole pockets in iron pnictides. More interestingly, recent inelastic neutron scattering measurements have revealed that the spin fluctuations in electron-doped iron selenide superconductor Li$_{1-x}$Fe$_x$ODFeSe present a twisted dispersion \cite{PanNC}, a dramatic deviation from the conventional spin wave-like dispersion typically observed in iron pnictides \cite{DaiRev,CaFeAs_SW}. This raises intriguing questions about the magnetic state from which these excitations originate, and the potential impact of electron doping on spin excitations.

In contrast to the typical method of achieving superconductivity in iron pnictides through chemical substitution into non-superconducting parent compounds, Li$_{1-x}$Fe$_x$ODFeSe achieves superconductivity through a distinct mechanism involving the intercalation of Li$_{1-x}$Fe$_x$OD between FeSe layers \cite{XianhuiNM,LiFeO_41K}. This intercalation process not only introduces carrier doping but also enhances the two-dimensional nature of the system. Both of these effects are crucial factors that could potentially play a pivotal role in enhancing superconductivity, as a record high critical temperature of around 65 K has been observed in mono-layer FeSe thin films \cite{FeSe_film_Tc1,FS_FeSe1,FS_FeSe2,FeSe_film_Tc2}.

The as-grown Li$_{1-x}$Fe$_x$ODFeSe compound typically exhibits superconductivity with a critical temperature (\textit{T}$_c$) of 41 K. Initially, ion gating was used to gradually modify the \textit{T}$_c$ of Li$_{1-x}$Fe$_x$ODFeSe, ultimately leading to the emergence of an insulating state as \textit{T}$_c$ diminished \cite{Gating}. However, ion-gated materials posed challenges for neutron scattering measurements, limiting our ability to explore the evolution of spin fluctuations. Recently, a significant breakthrough has been achieved by researchers who managed to induce the insulating state in Li$_{1-x}$Fe$_x$ODFeSe through precise modifications in the synthesis procedure \cite{LiFeOIns1,LiFeOIns2,LiFeOIns3,LiFeONPD}. This breakthrough opens an exciting and alternative avenue for further investigation into the material's properties and behavior.

\begin{figure}[ht!]
	\begin{center}
		\includegraphics[width=0.45\textwidth]{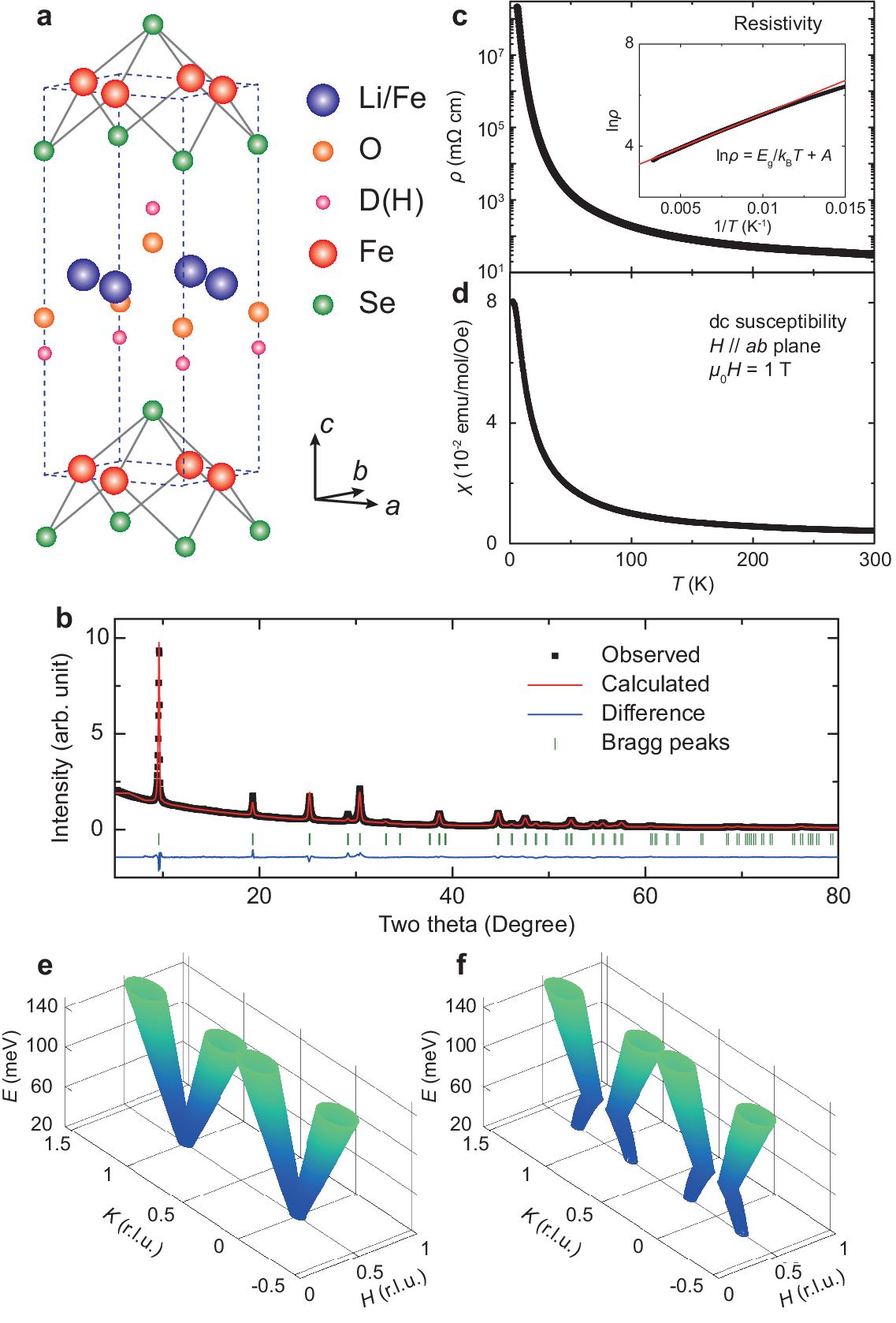}
	\end{center}
	\caption{Characterizations of insulating Li$_{1-x}$Fe$_x$ODFeSe compound and comparison of schematic diagrams of spin excitation spectra in insulating and superconducting Li$_{1-x}$Fe$_x$ODFeSe. (\textbf{a}) The schematics of crystal structure of Li$_{1-x}$Fe$_x$ODFeSe \cite{XianhuiNM}. (\textbf{b}) The X-ray powder diffraction pattern and Rietveld refinement conducted on the insulating phase of Li$_{1-x}$Fe$_x$ODFeSe, using ground single crystals. (\textbf{c}) In-plane resistivity measurement on the insulating phase of Li$_{1-x}$Fe$_x$ODFeSe single crystal. Four-probe method is employed. The inset shows a linear fitting of ln$\rho$-1/\textit{T} plot above 100 K. (\textbf{d}) Direct current (dc) magnetic susceptibility of the insulating Li$_{1-x}$Fe$_x$ODFeSe single crystal. External field $\mu$$_0$\textit{H} = 1 T is applied along \textit{ab} plane. (\textbf{e}) Schematic image of V-shaped spin excitation spectrum in insulating Li$_{1-x}$Fe$_x$ODFeSe. (\textbf{f}) Schematic of the hourglass spin excitation spectrum in superconducting Li$_{1-x}$Fe$_x$ODFeSe \cite{PanNC}. In (\textbf{e}) and (\textbf{f}), only excitations along the (0.5, \textit{K}) direction are shown for clarity.}
	\label{fig:fig1}
\end{figure}

In this Letter, we conducted an in-depth investigation into spin excitations within insulating Li$_{1-x}$Fe$_x$ODFeSe single crystals, employing neutron scattering techniques \cite{Supplement} (see also references \cite{XiaoliDong,Polarize,YFG} therein). Insulating Li$_{1-x}$Fe$_x$ODFeSe crystallizes in the P4/nmm space group, where FeSe and Li$_{1-x}$Fe$_x$OD layers alternate along the \textit{c} direction, as depicted in Fig.~\ref{fig:fig1}(a). The structural parameters were determined via Rietveld refinement of X-ray diffraction measurements conducted on ground single crystals (Fig.~\ref{fig:fig1}(b)), yielding lattice constants of \textit{a} = \textit{b} = 3.817 {\AA} and \textit{c} = 9.157 {\AA}. It's noteworthy that in comparison to optimally doped Li$_{1-x}$Fe$_x$ODFeSe \cite{PanNC,Boothroyd}, we observed lattice expansion within the \textit{ab} plane and simultaneous contraction along the \textit{c} axis in insulating Li$_{1-x}$Fe$_x$ODFeSe. This evolution in lattice parameters is proposed to stem from the reducing iron concentrations, effectively translating into a reduction in electron doping levels, as previously discussed in Refs. \cite{LiFeOIns1,LiFeOIns2,LiFeONPD,LiFeOIns3}. 

\begin{figure}[ht!]
	\begin{center}
		\includegraphics[width=0.4\textwidth]{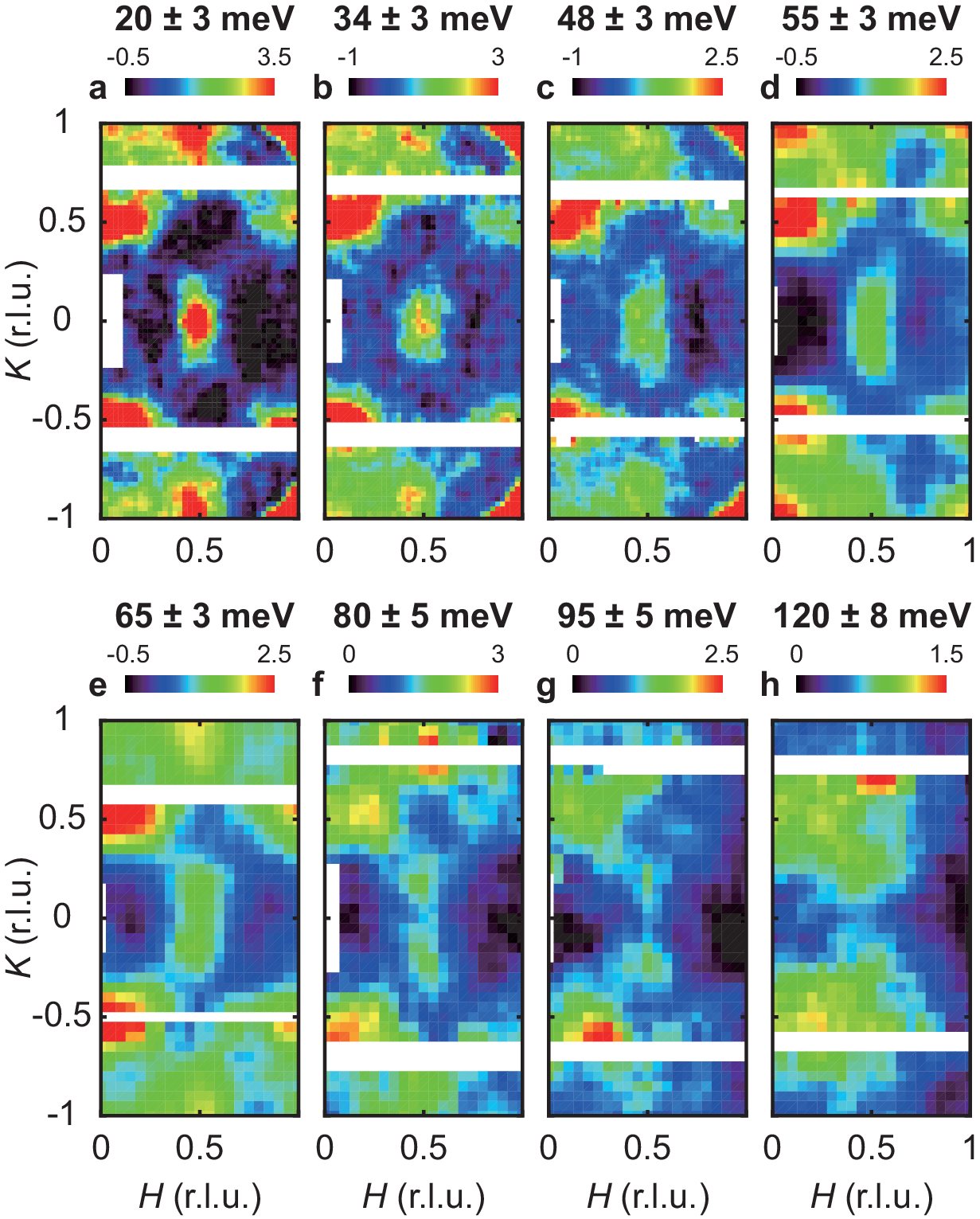}
	\end{center}
	\caption{(\textit{H}, \textit{K})-plane momentum dependence of the spin excitations in Li$_{1-x}$Fe$_x$ODFeSe insulator at 5 K with the energy transfers indicated. The background signals are subtracted as described in the main text. The measurements were conducted on ARCS, and the data analysis utilized the Mslice program. To improve plot quality, the data along the –\textit{H} direction was folded. The incident neutron energies used were 191.6 meV (\textbf{a}-\textbf{e}) and 310 meV (\textbf{f}-\textbf{h}). A vanadium run was performed for normalization process. The intensities are therefore presented in mbar sr$^{-1}$ meV$^{-1}$ f.u.$^{-1}$. }
	\label{fig:fig2}
\end{figure}

\begin{figure}[ht!]
	\begin{center}
		\includegraphics[width=0.45\textwidth]{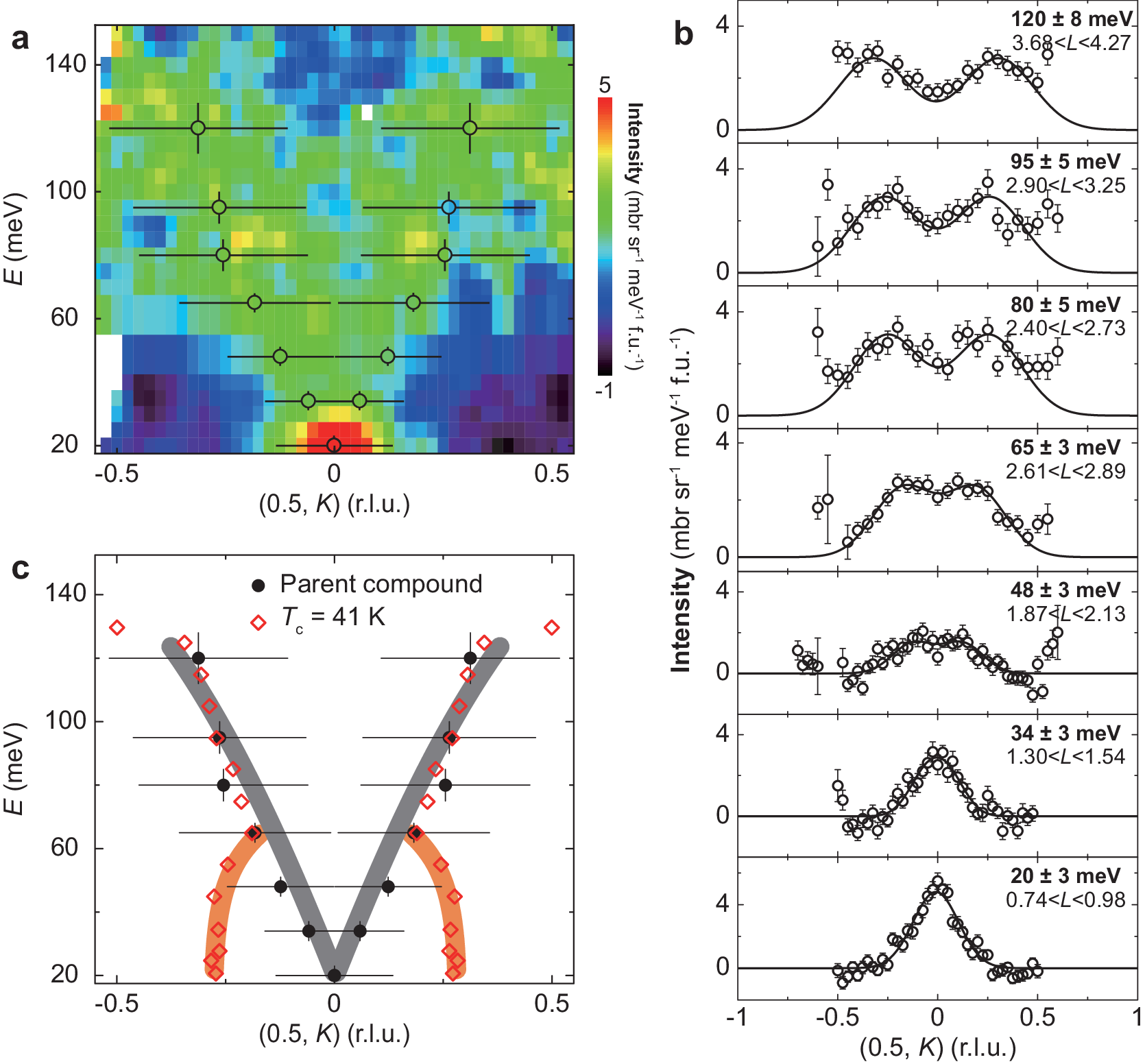}
	\end{center}
	\caption{Dispersion of the spin excitations in insulating Li$_{1-x}$Fe$_x$ODFeSe. (\textbf{a}) The background-subtracted spin excitations projected along the (0.5, \textit{K}) direction in insulating Li$_{1-x}$Fe$_x$ODFeSe at 5 K. The data was collected on ARCS, with energy ranges of 20 meV $\leq \textit{E} \leq$ 65 meV and 70 meV $\leq \textit{E} \leq$ 150 meV collected under the incident energies of 191.6 meV and 310 meV, respectively. The peak positions and full widths at half maximum (FWHM) of the spin excitation signals obtained from Gaussian fitting of constant energy cuts in (\textbf{b}) are represented by open circles and horizontal bars. The ranges of integrated energies are indicated by vertical bars. The isotropic Fe$^{2+}$ form factor has been corrected. (\textbf{b}) Constant-energy cuts of the stripe-type spin fluctuations along the (0.5, \textit{K}) direction at 5 K, with background signals eliminated and Fe$^{2+}$ form factor corrected. The energy/\textit{L} integration intervals were indicated. The fitting results of data using a Gaussian profile are illustrated by solid lines. The error bars correspond to one standard deviation. (\textbf{c}) The dispersion along the (0.5, \textit{K}) direction is overlaid for parent (black dots) and optimal-doped (red open diamonds) Li$_{1-x}$Fe$_x$ODFeSe. The gray and orange lines serve as guides to the eye. The data of optimal-doped sample is taken from Ref. \cite{PanNC}. }
	\label{fig:fig3}
\end{figure}

The resistivity of our single crystals exhibits insulating behavior over a wide temperature range, extending from 300 K down to 2 K, as illustrated in Fig.~\ref{fig:fig1}(c). We estimated the band gap to be approximately 20 meV by fitting the resistivity data above 100 K to the activation energy equation $\rho = A \exp{({E_g}/{k_B T})}$. Furthermore, the temperature-dependent behavior of the direct current (dc) magnetic susceptibility follows a Curie-Weiss-like behavior, as depicted in Fig.~\ref{fig:fig1}(d), and no discernible indications of long-range magnetic order were observed. These results are in general alignment with prior reports \cite{LiFeONPD,LiFeOIns3}. 

In Fig.~\ref{fig:fig2}, we present the momentum dependence of spin excitations in insulating Li$_{1-x}$Fe$_x$ODFeSe within the (\textit{H}, \textit{K}) plane at various energies (1-Fe unit cell). To enhance the clarity of the data, we subtracted the background signals dependent on $\lvert\textbf{Q}\rvert$, primarily arising from phonon excitations in the aluminum holder, for energy values below aluminum phonon-cutoff energy of 50 meV. Above 50 meV, we have adopted a constant background at each energy, following the approach outlined in Refs. \cite{PanNC,WangNC,TamPRB}. At lower energies, the spin excitations center around the stripe wave vector ($\pi,0$) and equivalent positions, forming an ellipse with long axis parallel to the transverse direction of reduced \textbf{q} (Fig.~\ref{fig:fig2}(a)). The elliptical pattern gradually elongates along its long axis with increasing energy (Fig.~\ref{fig:fig2}(b-d)). This behavior is in contrast to heavily electron doped Li$_{1-x}$Fe$_x$ODFeSe, where the low energy spin excitations exhibit a ring shaped incommensurate resonance mode surrounding ($\pi,\pi$) \cite{PanNC}. Above 65 meV, two distinct peaks become resolvable, and these split peaks continue to shift outward with increasing energy (Fig.~\ref{fig:fig2}(e-h)).

To elucidate the detailed dispersion of the excitations in insulating Li$_{1-x}$Fe$_x$ODFeSe, we present the excitation spectra in the \textit{E}-\textbf{Q} space. As depicted in Fig.~\ref{fig:fig3}(a), the spin excitations emanate from the stripe wavevector, progressively splitting into two branches as energy increases, eventually approaching the zone boundary at approximately 150 meV. For a quantitative analysis of the \textit{E}-\textbf{Q} relation, we conduct fitting procedures on the constant energy cuts using a Gaussian profile, as demonstrated in Fig.~\ref{fig:fig3}(b). We observe a single peak at around 20 meV, which progressively broadens as energy increases. With further energy increase, this peak bifurcates into two distinct peaks, resulting in a V-shaped dispersion pattern that is consistent with the contour plot depicted in Fig.~\ref{fig:fig3}(a).
 
The V-shaped spin excitations emanating from the ($\pi, 0$) point in insulating Li$_{1-x}$Fe$_x$ODFeSe closely resemble the spin waves observed in the antiferromagnetically ordered parent compounds of iron pnictide superconductors \cite{DaiRev,CaFeAs_SW}. These excitations primarily stem from the localized magnetic moments within the system, although the influence of itinerant electrons cannot be completely discounted, given the relatively modest energy gap observed in insulating Li$_{1-x}$Fe$_x$ODFeSe. Our estimation reveals that the spin wave velocity in insulating Li$_{1-x}$Fe$_x$ODFeSe is approximately $50\%$ lower than that in 122 iron pnictides \cite{CaFeAs_SW,BaFeAs_SW}. It is interesting to note that the Se height above the Fe plane is approximately 1.45 {\AA} in  Li$_{1-x}$Fe$_x$ODFeSe, which is greater than the pnictogen height in iron pnictides \cite{NaFeAs_SW,TamPRB,BaFeAsP}. This observation aligns with the general trend that a larger pnictogen height tends to enhance electronic correlations and reduce the spin wave velocity \cite{NaFeAs_SW,TamPRB,BaFeAsP,Zhiping1,Zhiping2}.

\begin{figure}[ht!]
	\begin{center}
		\includegraphics[width=0.4\textwidth]{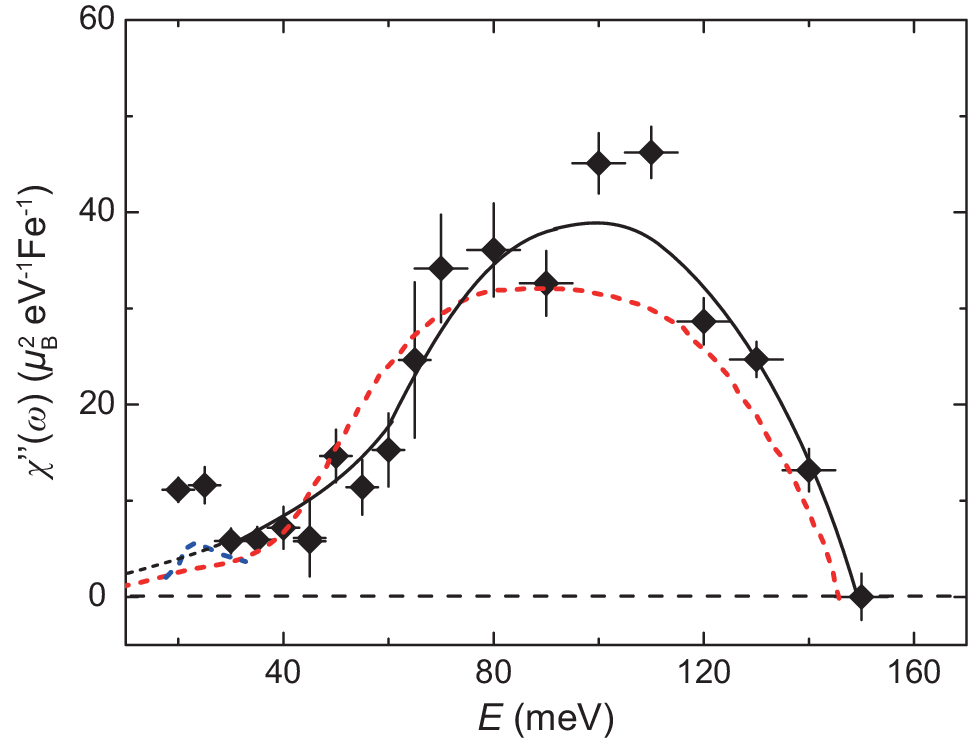}
	\end{center}
	\caption{Momentum-integrated local susceptibility \textit{$\chi$}''(\textit{$\omega$}) of the spin excitations in insulating Li$_{1-x}$Fe$_x$ODFeSe. The solid line is a guide to the eye. The horizontal bars indicate the integrating range of energies, and vertical bars represent error bars corresponding to one standard deviation. The data with 20 meV $\leq \textit{E} \leq 65$ meV and 70 meV $\leq \textit{E} \leq$ 150 meV was collected on ARCS, using the incident energies of 191.6 meV and 310 meV respectively. The red dashed line shows \textit{$\chi$}''(\textit{$\omega$}) in superconducting Li$_{1-x}$Fe$_x$ODFeSe and the blue dash represents for the spin resonance mode, adopted from Ref. \cite{PanNC}. These results suggest that superconductivity under electron doping does not significantly change the total fluctuation moment, similar to observations in iron pnictides \cite{122_moment,LuoBaNi}.}
	\label{fig:fig4}
\end{figure}  

To gain a more profound understanding of the magnetism in insulating Li$_{1-x}$Fe$_x$ODFeSe, we present local susceptibility as a function of energy, which provides an estimate of the fluctuating moment of Fe ions (Fig.~\ref{fig:fig4}). The local susceptibility was obtained by averaging the spin spectral weight over the Brillouin zone, followed by the Bose factor correction. The resulting $\chi$''($\omega$) profile reveals a broad hump centered around 100 meV, with an upper limit extending to approximately 150 meV.
This value is smaller than the bandwidth of iron pnictides \cite{DaiRev,122_moment, MengWang,LuoBaNi}, consistent with the softened spin wave velocity observed in Li$_{1-x}$Fe$_x$ODFeSe compared to iron pnicitides.

By summing $\chi$''($\omega$) across the energy range, we determined the fluctuating moment to be (2.92$\pm$0.38) $\mu$$_B^2$/Fe, corresponding to an effective spin (\textit{S}) value of 1/2. The fluctuating moment in the insulating Li$_{1-x}$Fe$_x$ODFeSe system appears to be smaller in magnitude when compared to that of FeSe and 122 iron pnictides \cite{WangNC,122_moment,112_Luo}. This observation is somewhat surprising, particularly considering the insulating ground state in Li$_{1-x}$Fe$_x$ODFeSe. One plausible explanation for this difference could be linked to the relatively large interlayer spacing induced by intercalation, which enhances fluctuations and subsequently leads to a reduction in the intensity of spin excitations. This phenomenon draws parallels with the behavior seen in quasi-two-dimensional cuprate superconductors, characterized by the presence of 'missing' spin-excitation spectral weight \cite{MissingSpec,CuprateRMP}.

It is informative to compare the spin excitations of the insulating and heavily electron-doped superconducting Li$_{1-x}$Fe$_x$ODFeSe (Fig.~\ref{fig:fig3}(c)). It is evident that the dispersion behavior of spin excitations in these two compounds above 65 meV are quite similar. A divergence becomes apparent at lower energy ranges, giving rise to distinct V-shaped and hourglass spectra in the insulating and superconducting states, respectively.

More specifically, electron doping predominantly influences low-energy spin excitations, causing a shift from the ($\pi,0$) point to an incommensurate position toward the ($\pi,\pi$) direction below 65 meV. These observations suggest a strong link between low-energy spin excitations and itinerant electrons, with a marked dependence on the level of doping. Conversely, high-energy spin excitations seem relatively unaffected by doping, suggesting a relationship with local moments akin to those in the insulating parent compound.

Most notably, the transition from a V-shaped dispersion to an hourglass-like pattern as doping progresses from the insulating to the superconducting state bears a striking resemblance to the behavior observed in hole-doped cuprates (Fig.~\ref{fig:fig1}(e),(f)) \cite{Hourglass1,Hourglass2,Hourglass3}. In cuprates, the extensively studied hourglass dispersion has led to various proposed models for its origin, including the Fermi surface nesting hypothesis \cite{Hourglass1,Bulut,QMSi,Littlewood,Norman,Hussey} and the induction of dynamic spin stripes due to hole doping \cite{Hourglass2,stripe1,stripe2}.

The similarity between Li$_{1-x}$Fe$_x$ODFeSe and cuprates implies a shared underlying mechanism. If this hypothesis holds true, it implies that the simultaneous presence of two distinct types of spin excitations is a recurring characteristic among high-temperature superconductors. In heavily electron-doped Li$_{1-x}$Fe$_x$ODFeSe and hole-doped cuprates, the Fermi surface nesting possesses different wavevectors from local moment antiferromagnetic interactions. The intersection of two types of excitations at a specific energy gives rise to the distinctive twisted hourglass dispersion. Conversely, carrier doping induces broadening of spin wave excitations when Fermi surface nesting aligns with a similar wavevector as the parent compound's antiferromagnetic interactions. These situations are illustrated in the case of iron pnictides \cite{DaiRev,122_moment,LuoBaNi} and electron-doped cuprates \cite{Wilson_PLCCO,Jun_PLCCO,ntypeCu}.

The elucidation of the nature of spin excitations and their doping dependence has the potential to shed light on the fundamental mechanisms underlying superconductivity. In the case of cuprates, the \textit{d}-wave pairing symmetry has gained substantial support and is thought to be intricately linked with the characteristic hourglass-shaped spin excitations \cite{CuprateRMP,Eschrig}. When considering Li$_{1-x}$Fe$_x$ODFeSe, the \textit{s}$_{\pm}$-pairing mechanism, typically observed in iron pnictides and reliant on the presence of both electron and hole pockets, cannot be readily applied due to the absence of hole pockets in this system.

Therefore, it is imperative to embark on further investigations to determine whether the hourglass-shaped spin excitations also align with a \textit{d}-wave pairing scenario in Li$_{1-x}$Fe$_x$ODFeSe, potentially associated with the electron pockets at zone corners \cite{dwave_FeSe1,dwave_FeSe2,dwave_FeSe3,dwave_FeSe4}. Additionally, delving into the behavior of spin excitations in the intermediate doping regime holds particular interest, as it could offer crucial insights into how the hourglass dispersion and superconductivity evolve with doping. Furthermore, it is interesting to explore the possibility of other competing orders \cite{Gating,ChargeOrder,K_dose} or pseudogap phenomena \cite{pseudogap1,pseudogap2,pseudogap3} in these intermediate doping regimes. Such phenomena are commonly observed in cuprates and could shed valuable light on the complex interplay of electronic states and correlations in these materials.

In summary, we present detailed neutron scattering measurements that explore spin correlations in insulating Li$_{1-x}$Fe$_x$ODFeSe. We observed a smaller total fluctuating moment in this material compared to other iron based superconductors. This difference could be due to extended interlayer distances resulting from intercalation, which boosts fluctuations while reducing spin excitation intensity. We also discovered a V-shaped spin wave-like excitation pattern in insulating Li$_{1-x}$Fe$_x$ODFeSe, where the spin wave velocity is $50\%$ lower than in 122 iron pnictides, likely a result of stronger correlations reducing magnetic excitation bandwidth. As doping transforms the insulating state into the superconducting state, it transitions from a V-shaped to an hourglass-like dispersion, a behavior reminiscent of what is observed in hole-doped cuprates. This intriguing parallel suggests the possibility of a shared underlying mechanism governing magnetism and superconductivity across these distinct systems. Our findings not only enrich the understanding of spin dynamics in iron-based materials but also pave the way for future investigations into the universal aspects of high-temperature superconductors and related materials.

\begin{center}
	ACKNOWLEDGEMENTS
\end{center}

This work was supported by National Key R$\&$D Program of China (2022YFA1403202), the Key Program of National Natural Science Foundation of China (Grant No. 12234006), the Shanghai Municipal Science and Technology Major Project (Grant No. 2019SHZDZX01),  the Innovation Program for Quantum Science and Technology (Grant No. 2024ZD0300103), and the Large Scientific Facility Open Subject of Songshan Lake Laboratory (Grant No. DG23313511). H. W. was supported by the Youth Foundation of the National Natural Science Foundation of China (Grant No. 12204108). B. P. is supported by the Natural Science Foundation of Shandong Province (Grant No. ZR2020YQ03). Y. F. is supported by the National Key R$\&$D Program of China (Grand No. 2022YFA1604104) and the Fundamental and Applied Fundamental Research Grant of Guangdong Province (Grant No. 2021B1515120015). A portion of this research used resources at the Spallation Neutron Source, a DOE Office of Science User Facility operated by the Oak Ridge National Laboratory. 

~\\
\noindent
$^\ast$ zhaoj@fudan.edu.cn

\noindent	
$^\dagger$ These authors contributed equally to this work.

\normalem


\begin{thebibliography}{67}%
	\makeatletter
	\providecommand \@ifxundefined [1]{%
		\@ifx{#1\undefined}
	}%
	\providecommand \@ifnum [1]{%
		\ifnum #1\expandafter \@firstoftwo
		\else \expandafter \@secondoftwo
		\fi
	}%
	\providecommand \@ifx [1]{%
		\ifx #1\expandafter \@firstoftwo
		\else \expandafter \@secondoftwo
		\fi
	}%
	\providecommand \natexlab [1]{#1}%
	\providecommand \enquote  [1]{``#1''}%
	\providecommand \bibnamefont  [1]{#1}%
	\providecommand \bibfnamefont [1]{#1}%
	\providecommand \citenamefont [1]{#1}%
	\providecommand \href@noop [0]{\@secondoftwo}%
	\providecommand \href [0]{\begingroup \@sanitize@url \@href}%
	\providecommand \@href[1]{\@@startlink{#1}\@@href}%
	\providecommand \@@href[1]{\endgroup#1\@@endlink}%
	\providecommand \@sanitize@url [0]{\catcode `\\12\catcode `\$12\catcode
		`\&12\catcode `\#12\catcode `\^12\catcode `\_12\catcode `\%12\relax}%
	\providecommand \@@startlink[1]{}%
	\providecommand \@@endlink[0]{}%
	\providecommand \url  [0]{\begingroup\@sanitize@url \@url }%
	\providecommand \@url [1]{\endgroup\@href {#1}{\urlprefix }}%
	\providecommand \urlprefix  [0]{URL }%
	\providecommand \Eprint [0]{\href }%
	\providecommand \doibase [0]{https://doi.org/}%
	\providecommand \selectlanguage [0]{\@gobble}%
	\providecommand \bibinfo  [0]{\@secondoftwo}%
	\providecommand \bibfield  [0]{\@secondoftwo}%
	\providecommand \translation [1]{[#1]}%
	\providecommand \BibitemOpen [0]{}%
	\providecommand \bibitemStop [0]{}%
	\providecommand \bibitemNoStop [0]{.\EOS\space}%
	\providecommand \EOS [0]{\spacefactor3000\relax}%
	\providecommand \BibitemShut  [1]{\csname bibitem#1\endcsname}%
	\let\auto@bib@innerbib\@empty
	%</preamble>
	\bibitem [{\citenamefont {Kamihara}\ \emph {et~al.}(2008)\citenamefont
		{Kamihara}, \citenamefont {Watanabe}, \citenamefont {Hirano},\ and\
		\citenamefont {Hosono}}]{Hosono}%
	\BibitemOpen
	\bibfield  {author} {\bibinfo {author} {\bibfnamefont {Y.}~\bibnamefont
			{Kamihara}}, \bibinfo {author} {\bibfnamefont {T.}~\bibnamefont {Watanabe}},
		\bibinfo {author} {\bibfnamefont {M.}~\bibnamefont {Hirano}},\ and\ \bibinfo
		{author} {\bibfnamefont {H.}~\bibnamefont {Hosono}},\ }\bibfield  {title}
	{\bibinfo {title} {Iron-based layered superconductor
			{La[O$_{1-\textit{x}}$F$_\textit{x}$]FeAs} (\textit{x} = 0.05-0.12) with
			\textit{{T}}$_c$ = 26 {K}},\ }\href {https://doi.org/10.1021/ja800073m}
	{\bibfield  {journal} {\bibinfo  {journal} {J. Am. Chem. Soc.}\ }\textbf
		{\bibinfo {volume} {130}},\ \bibinfo {pages} {3296} (\bibinfo {year}
		{2008})}\BibitemShut {NoStop}%
	\bibitem [{\citenamefont {Dai}(2015)}]{DaiRev}%
	\BibitemOpen
	\bibfield  {author} {\bibinfo {author} {\bibfnamefont {P.~C.}\ \bibnamefont
			{Dai}},\ }\bibfield  {title} {\bibinfo {title} {Antiferromagnetic order and
			spin dynamics in iron-based superconductors},\ }\href
	{https://doi.org/10.1103/RevModPhys.87.855} {\bibfield  {journal} {\bibinfo
			{journal} {Rev. Mod. Phys.}\ }\textbf {\bibinfo {volume} {87}},\ \bibinfo
		{pages} {855} (\bibinfo {year} {2015})}\BibitemShut {NoStop}%
	\bibitem [{\citenamefont {Scalapino}(2012)}]{ScalapinoRev}%
	\BibitemOpen
	\bibfield  {author} {\bibinfo {author} {\bibfnamefont {D.~J.}\ \bibnamefont
			{Scalapino}},\ }\bibfield  {title} {\bibinfo {title} {A common thread: the
			pairing interaction for the unconventional superconductors},\ }\href
	{https://doi.org/10.1103/RevModPhys.84.1383} {\bibfield  {journal} {\bibinfo
			{journal} {Rev. Mod. Phys.}\ }\textbf {\bibinfo {volume} {84}},\ \bibinfo
		{pages} {1383} (\bibinfo {year} {2012})}\BibitemShut {NoStop}%
	\bibitem [{\citenamefont {Guo}\ \emph {et~al.}(2010)\citenamefont {Guo},
		\citenamefont {Jin}, \citenamefont {Wang}, \citenamefont {Wang},
		\citenamefont {Zhu}, \citenamefont {Zhou}, \citenamefont {He},\ and\
		\citenamefont {Chen}}]{KFe2Se2}%
	\BibitemOpen
	\bibfield  {author} {\bibinfo {author} {\bibfnamefont {J.}~\bibnamefont
			{Guo}}, \bibinfo {author} {\bibfnamefont {S.}~\bibnamefont {Jin}}, \bibinfo
		{author} {\bibfnamefont {G.}~\bibnamefont {Wang}}, \bibinfo {author}
		{\bibfnamefont {S.}~\bibnamefont {Wang}}, \bibinfo {author} {\bibfnamefont
			{K.}~\bibnamefont {Zhu}}, \bibinfo {author} {\bibfnamefont {T.}~\bibnamefont
			{Zhou}}, \bibinfo {author} {\bibfnamefont {M.}~\bibnamefont {He}},\ and\
		\bibinfo {author} {\bibfnamefont {X.}~\bibnamefont {Chen}},\ }\bibfield
	{title} {\bibinfo {title} {Superconductivity in the iron selenide
			{K$_\textit{x}$Fe$_2$Se$_2$} (0$\leq$\textit{x}$\leq$1.0)},\ }\href
	{https://doi.org/10.1103/PhysRevB.82.180520} {\bibfield  {journal} {\bibinfo
			{journal} {Phys. Rev. B}\ }\textbf {\bibinfo {volume} {82}},\ \bibinfo
		{pages} {180520} (\bibinfo {year} {2010})}\BibitemShut {NoStop}%
	\bibitem [{\citenamefont {Wang}\ \emph {et~al.}(2012)\citenamefont {Wang},
		\citenamefont {Li}, \citenamefont {Zhang}, \citenamefont {Zhang},
		\citenamefont {Zhang}, \citenamefont {Li}, \citenamefont {Ding},
		\citenamefont {Ou}, \citenamefont {Deng}, \citenamefont {Chang},
		\citenamefont {Wen}, \citenamefont {Song}, \citenamefont {He}, \citenamefont
		{Jia}, \citenamefont {Ji}, \citenamefont {Wang}, \citenamefont {Wang},
		\citenamefont {Chen}, \citenamefont {Ma},\ and\ \citenamefont
		{Xue}}]{FeSe_film_Tc1}%
	\BibitemOpen
	\bibfield  {author} {\bibinfo {author} {\bibfnamefont {Q.-Y.}\ \bibnamefont
			{Wang}}, \bibinfo {author} {\bibfnamefont {Z.}~\bibnamefont {Li}}, \bibinfo
		{author} {\bibfnamefont {W.-H.}\ \bibnamefont {Zhang}}, \bibinfo {author}
		{\bibfnamefont {Z.-C.}\ \bibnamefont {Zhang}}, \bibinfo {author}
		{\bibfnamefont {J.-S.}\ \bibnamefont {Zhang}}, \bibinfo {author}
		{\bibfnamefont {W.}~\bibnamefont {Li}}, \bibinfo {author} {\bibfnamefont
			{H.}~\bibnamefont {Ding}}, \bibinfo {author} {\bibfnamefont {Y.-B.}\
			\bibnamefont {Ou}}, \bibinfo {author} {\bibfnamefont {P.}~\bibnamefont
			{Deng}}, \bibinfo {author} {\bibfnamefont {K.}~\bibnamefont {Chang}},
		\bibinfo {author} {\bibfnamefont {J.}~\bibnamefont {Wen}}, \bibinfo {author}
		{\bibfnamefont {C.-L.}\ \bibnamefont {Song}}, \bibinfo {author}
		{\bibfnamefont {K.}~\bibnamefont {He}}, \bibinfo {author} {\bibfnamefont
			{J.-F.}\ \bibnamefont {Jia}}, \bibinfo {author} {\bibfnamefont {S.-H.}\
			\bibnamefont {Ji}}, \bibinfo {author} {\bibfnamefont {Y.-Y.}\ \bibnamefont
			{Wang}}, \bibinfo {author} {\bibfnamefont {L.-L.}\ \bibnamefont {Wang}},
		\bibinfo {author} {\bibfnamefont {X.}~\bibnamefont {Chen}}, \bibinfo {author}
		{\bibfnamefont {X.-C.}\ \bibnamefont {Ma}},\ and\ \bibinfo {author}
		{\bibfnamefont {Q.-K.}\ \bibnamefont {Xue}},\ }\bibfield  {title} {\bibinfo
		{title} {Interface induced high temperature superconductivity in single
			unit-cell {FeSe} films on {SrTiO$_3$}},\ }\href
	{https://doi.org/10.1088/0256-307X/29/3/037402} {\bibfield  {journal}
		{\bibinfo  {journal} {Chin. Phys. Lett.}\ }\textbf {\bibinfo {volume} {29}},\
		\bibinfo {pages} {037402} (\bibinfo {year} {2012})}\BibitemShut {NoStop}%
	\bibitem [{\citenamefont {Lu}\ \emph {et~al.}(2015)\citenamefont {Lu},
		\citenamefont {Wang}, \citenamefont {Wu}, \citenamefont {Wu}, \citenamefont
		{Zhao}, \citenamefont {Zeng}, \citenamefont {Luo}, \citenamefont {Wu},
		\citenamefont {Bao}, \citenamefont {Zhang}, \citenamefont {Huang},
		\citenamefont {Huang},\ and\ \citenamefont {Chen}}]{XianhuiNM}%
	\BibitemOpen
	\bibfield  {author} {\bibinfo {author} {\bibfnamefont {X.~F.}\ \bibnamefont
			{Lu}}, \bibinfo {author} {\bibfnamefont {N.~Z.}\ \bibnamefont {Wang}},
		\bibinfo {author} {\bibfnamefont {H.}~\bibnamefont {Wu}}, \bibinfo {author}
		{\bibfnamefont {Y.~P.}\ \bibnamefont {Wu}}, \bibinfo {author} {\bibfnamefont
			{D.}~\bibnamefont {Zhao}}, \bibinfo {author} {\bibfnamefont {X.~Z.}\
			\bibnamefont {Zeng}}, \bibinfo {author} {\bibfnamefont {X.~G.}\ \bibnamefont
			{Luo}}, \bibinfo {author} {\bibfnamefont {T.}~\bibnamefont {Wu}}, \bibinfo
		{author} {\bibfnamefont {W.}~\bibnamefont {Bao}}, \bibinfo {author}
		{\bibfnamefont {G.~H.}\ \bibnamefont {Zhang}}, \bibinfo {author}
		{\bibfnamefont {F.~Q.}\ \bibnamefont {Huang}}, \bibinfo {author}
		{\bibfnamefont {Q.~Z.}\ \bibnamefont {Huang}},\ and\ \bibinfo {author}
		{\bibfnamefont {X.~H.}\ \bibnamefont {Chen}},\ }\bibfield  {title} {\bibinfo
		{title} {Coexistence of superconductivity and antiferromagnetism in
			{(Li$_{0.8}$Fe$_{0.2}$)OHFeSe}},\ }\href {https://doi.org/10.1038/nmat4155}
	{\bibfield  {journal} {\bibinfo  {journal} {Nat. Mater.}\ }\textbf {\bibinfo
			{volume} {14}},\ \bibinfo {pages} {325} (\bibinfo {year} {2015})}\BibitemShut
	{NoStop}%
	\bibitem [{\citenamefont {Pachmayr}\ \emph {et~al.}(2015)\citenamefont
		{Pachmayr}, \citenamefont {Nitsche}, \citenamefont {Luetkens}, \citenamefont
		{Kamusella}, \citenamefont {Br$\ddot{u}$ckner}, \citenamefont {Sarkar},
		\citenamefont {Klauss},\ and\ \citenamefont {Johrendt}}]{LiFeO_41K}%
	\BibitemOpen
	\bibfield  {author} {\bibinfo {author} {\bibfnamefont {U.}~\bibnamefont
			{Pachmayr}}, \bibinfo {author} {\bibfnamefont {F.}~\bibnamefont {Nitsche}},
		\bibinfo {author} {\bibfnamefont {H.}~\bibnamefont {Luetkens}}, \bibinfo
		{author} {\bibfnamefont {S.}~\bibnamefont {Kamusella}}, \bibinfo {author}
		{\bibfnamefont {F.}~\bibnamefont {Br$\ddot{u}$ckner}}, \bibinfo {author}
		{\bibfnamefont {R.}~\bibnamefont {Sarkar}}, \bibinfo {author} {\bibfnamefont
			{H.-H.}\ \bibnamefont {Klauss}},\ and\ \bibinfo {author} {\bibfnamefont
			{D.}~\bibnamefont {Johrendt}},\ }\bibfield  {title} {\bibinfo {title}
		{Coexistence of 3\textit{d}-ferromagnetism and superconductivity in
			{[(Li$_{1-\textit{x}}$Fe$_\textit{x}$)OH](Fe$_{1-\textit{y}}$Li$_\textit{y}$)Se}},\
	}\href {https://doi.org/10.1002/anie.201407756} {\bibfield  {journal}
		{\bibinfo  {journal} {Angew. Chem. Int. Ed.}\ }\textbf {\bibinfo {volume}
			{54}},\ \bibinfo {pages} {293} (\bibinfo {year} {2015})}\BibitemShut
	{NoStop}%
	\bibitem [{\citenamefont {Zhang}\ \emph {et~al.}(2011)\citenamefont {Zhang},
		\citenamefont {Yang}, \citenamefont {Xu}, \citenamefont {Ye}, \citenamefont
		{Chen}, \citenamefont {He}, \citenamefont {Xu}, \citenamefont {Jiang},
		\citenamefont {Xie}, \citenamefont {Ying}, \citenamefont {F.Wang},
		\citenamefont {Chen}, \citenamefont {Hu}, \citenamefont {Matsunami},
		\citenamefont {Kimura},\ and\ \citenamefont {Feng}}]{FS_AFeSe1}%
	\BibitemOpen
	\bibfield  {author} {\bibinfo {author} {\bibfnamefont {Y.}~\bibnamefont
			{Zhang}}, \bibinfo {author} {\bibfnamefont {L.~X.}\ \bibnamefont {Yang}},
		\bibinfo {author} {\bibfnamefont {M.}~\bibnamefont {Xu}}, \bibinfo {author}
		{\bibfnamefont {Z.~R.}\ \bibnamefont {Ye}}, \bibinfo {author} {\bibfnamefont
			{F.}~\bibnamefont {Chen}}, \bibinfo {author} {\bibfnamefont {C.}~\bibnamefont
			{He}}, \bibinfo {author} {\bibfnamefont {H.~C.}\ \bibnamefont {Xu}}, \bibinfo
		{author} {\bibfnamefont {J.}~\bibnamefont {Jiang}}, \bibinfo {author}
		{\bibfnamefont {B.~P.}\ \bibnamefont {Xie}}, \bibinfo {author} {\bibfnamefont
			{J.~J.}\ \bibnamefont {Ying}}, \bibinfo {author} {\bibfnamefont
			{X.}~\bibnamefont {F.Wang}}, \bibinfo {author} {\bibfnamefont {X.~H.}\
			\bibnamefont {Chen}}, \bibinfo {author} {\bibfnamefont {J.~P.}\ \bibnamefont
			{Hu}}, \bibinfo {author} {\bibfnamefont {M.}~\bibnamefont {Matsunami}},
		\bibinfo {author} {\bibfnamefont {S.}~\bibnamefont {Kimura}},\ and\ \bibinfo
		{author} {\bibfnamefont {D.~L.}\ \bibnamefont {Feng}},\ }\bibfield  {title}
	{\bibinfo {title} {Nodeless superconducting gap in
			{A$_\textit{x}$Fe$_2$Se$_2$ (A=K,Cs)} revealed by angle-resolved
			photoemission spectroscopy},\ }\href {https://doi.org/10.1038/NMAT2981}
	{\bibfield  {journal} {\bibinfo  {journal} {Nat. Mater.}\ }\textbf {\bibinfo
			{volume} {10}},\ \bibinfo {pages} {273} (\bibinfo {year} {2011})}\BibitemShut
	{NoStop}%
	\bibitem [{\citenamefont {Qian}\ \emph {et~al.}(2011)\citenamefont {Qian},
		\citenamefont {X.-P.Wang}, \citenamefont {Jin}, \citenamefont {Zhang},
		\citenamefont {Richard}, \citenamefont {Xu}, \citenamefont {Dai},
		\citenamefont {Fang}, \citenamefont {Guo}, \citenamefont {Chen},\ and\
		\citenamefont {Ding}}]{FS_AFeSe2}%
	\BibitemOpen
	\bibfield  {author} {\bibinfo {author} {\bibfnamefont {T.}~\bibnamefont
			{Qian}}, \bibinfo {author} {\bibnamefont {X.-P.Wang}}, \bibinfo {author}
		{\bibfnamefont {W.-C.}\ \bibnamefont {Jin}}, \bibinfo {author} {\bibfnamefont
			{P.}~\bibnamefont {Zhang}}, \bibinfo {author} {\bibfnamefont
			{P.}~\bibnamefont {Richard}}, \bibinfo {author} {\bibfnamefont
			{G.}~\bibnamefont {Xu}}, \bibinfo {author} {\bibfnamefont {X.}~\bibnamefont
			{Dai}}, \bibinfo {author} {\bibfnamefont {Z.}~\bibnamefont {Fang}}, \bibinfo
		{author} {\bibfnamefont {J.-G.}\ \bibnamefont {Guo}}, \bibinfo {author}
		{\bibfnamefont {X.-L.}\ \bibnamefont {Chen}},\ and\ \bibinfo {author}
		{\bibfnamefont {H.}~\bibnamefont {Ding}},\ }\bibfield  {title} {\bibinfo
		{title} {Absence of a holelike {Fermi} surface for the iron-based
			{K$_{0.8}$Fe$_{1.7}$Se$_2$} superconductor revealed by angle-resolved
			photoemission spectroscopy},\ }\href
	{https://doi.org/10.1103/PhysRevLett.106.187001} {\bibfield  {journal}
		{\bibinfo  {journal} {Phys. Rev. Lett.}\ }\textbf {\bibinfo {volume} {106}},\
		\bibinfo {pages} {187001} (\bibinfo {year} {2011})}\BibitemShut {NoStop}%
	\bibitem [{\citenamefont {Mou}\ \emph {et~al.}(2011)\citenamefont {Mou},
		\citenamefont {Liu}, \citenamefont {Jia}, \citenamefont {He}, \citenamefont
		{Peng}, \citenamefont {Zhao}, \citenamefont {Yu}, \citenamefont {Liu},
		\citenamefont {He}, \citenamefont {Dong}, \citenamefont {Zhang},
		\citenamefont {Wang}, \citenamefont {Dong}, \citenamefont {Fang},
		\citenamefont {Wang}, \citenamefont {Peng}, \citenamefont {Wang},
		\citenamefont {Zhang}, \citenamefont {Yang}, \citenamefont {Xu},
		\citenamefont {Chen},\ and\ \citenamefont {Zhou}}]{FS_AFeSe3}%
	\BibitemOpen
	\bibfield  {author} {\bibinfo {author} {\bibfnamefont {D.}~\bibnamefont
			{Mou}}, \bibinfo {author} {\bibfnamefont {S.}~\bibnamefont {Liu}}, \bibinfo
		{author} {\bibfnamefont {X.}~\bibnamefont {Jia}}, \bibinfo {author}
		{\bibfnamefont {J.}~\bibnamefont {He}}, \bibinfo {author} {\bibfnamefont
			{Y.}~\bibnamefont {Peng}}, \bibinfo {author} {\bibfnamefont {L.}~\bibnamefont
			{Zhao}}, \bibinfo {author} {\bibfnamefont {L.}~\bibnamefont {Yu}}, \bibinfo
		{author} {\bibfnamefont {G.}~\bibnamefont {Liu}}, \bibinfo {author}
		{\bibfnamefont {S.}~\bibnamefont {He}}, \bibinfo {author} {\bibfnamefont
			{X.}~\bibnamefont {Dong}}, \bibinfo {author} {\bibfnamefont {J.}~\bibnamefont
			{Zhang}}, \bibinfo {author} {\bibfnamefont {H.}~\bibnamefont {Wang}},
		\bibinfo {author} {\bibfnamefont {C.}~\bibnamefont {Dong}}, \bibinfo {author}
		{\bibfnamefont {M.}~\bibnamefont {Fang}}, \bibinfo {author} {\bibfnamefont
			{X.}~\bibnamefont {Wang}}, \bibinfo {author} {\bibfnamefont {Q.}~\bibnamefont
			{Peng}}, \bibinfo {author} {\bibfnamefont {Z.}~\bibnamefont {Wang}}, \bibinfo
		{author} {\bibfnamefont {S.}~\bibnamefont {Zhang}}, \bibinfo {author}
		{\bibfnamefont {F.}~\bibnamefont {Yang}}, \bibinfo {author} {\bibfnamefont
			{Z.}~\bibnamefont {Xu}}, \bibinfo {author} {\bibfnamefont {C.}~\bibnamefont
			{Chen}},\ and\ \bibinfo {author} {\bibfnamefont {X.~J.}\ \bibnamefont
			{Zhou}},\ }\bibfield  {title} {\bibinfo {title} {Distinct {Fermi} surface
			topology and nodeless superconducting gap in a
			{(Tl$_{0.58}$Rb$_{0.42}$)Fe$_{1.72}$Se$_2$} superconductor},\ }\href
	{https://doi.org/10.1103/PhysRevLett.106.107001} {\bibfield  {journal}
		{\bibinfo  {journal} {Phys. Rev. Lett.}\ }\textbf {\bibinfo {volume} {106}},\
		\bibinfo {pages} {107001} (\bibinfo {year} {2011})}\BibitemShut {NoStop}%
	\bibitem [{\citenamefont {Dagotto}(2013)}]{FS_AFeSe4}%
	\BibitemOpen
	\bibfield  {author} {\bibinfo {author} {\bibfnamefont {E.}~\bibnamefont
			{Dagotto}},\ }\bibfield  {title} {\bibinfo {title} {The unexpected properties
			of alkali metal iron selenide superconductors},\ }\href
	{https://doi.org/10.1103/RevModPhys.85.849} {\bibfield  {journal} {\bibinfo
			{journal} {Rev. Mod. Phys.}\ }\textbf {\bibinfo {volume} {85}},\ \bibinfo
		{pages} {849} (\bibinfo {year} {2013})}\BibitemShut {NoStop}%
	\bibitem [{\citenamefont {He}\ \emph {et~al.}(2013)\citenamefont {He},
		\citenamefont {He}, \citenamefont {Zhang}, \citenamefont {Zhao},
		\citenamefont {Liu}, \citenamefont {Liu}, \citenamefont {Mou}, \citenamefont
		{Ou}, \citenamefont {Wang}, \citenamefont {Li}, \citenamefont {Wang},
		\citenamefont {Peng}, \citenamefont {Liu}, \citenamefont {Chen},
		\citenamefont {Yu}, \citenamefont {Liu}, \citenamefont {Dong}, \citenamefont
		{Zhang}, \citenamefont {Chen}, \citenamefont {Xu}, \citenamefont {Chen},
		\citenamefont {Ma}, \citenamefont {Xue},\ and\ \citenamefont
		{Zhou}}]{FS_FeSe1}%
	\BibitemOpen
	\bibfield  {author} {\bibinfo {author} {\bibfnamefont {S.}~\bibnamefont
			{He}}, \bibinfo {author} {\bibfnamefont {J.}~\bibnamefont {He}}, \bibinfo
		{author} {\bibfnamefont {W.}~\bibnamefont {Zhang}}, \bibinfo {author}
		{\bibfnamefont {L.}~\bibnamefont {Zhao}}, \bibinfo {author} {\bibfnamefont
			{D.}~\bibnamefont {Liu}}, \bibinfo {author} {\bibfnamefont {X.}~\bibnamefont
			{Liu}}, \bibinfo {author} {\bibfnamefont {D.}~\bibnamefont {Mou}}, \bibinfo
		{author} {\bibfnamefont {Y.-B.}\ \bibnamefont {Ou}}, \bibinfo {author}
		{\bibfnamefont {Q.-Y.}\ \bibnamefont {Wang}}, \bibinfo {author}
		{\bibfnamefont {Z.}~\bibnamefont {Li}}, \bibinfo {author} {\bibfnamefont
			{L.}~\bibnamefont {Wang}}, \bibinfo {author} {\bibfnamefont {Y.}~\bibnamefont
			{Peng}}, \bibinfo {author} {\bibfnamefont {Y.}~\bibnamefont {Liu}}, \bibinfo
		{author} {\bibfnamefont {C.}~\bibnamefont {Chen}}, \bibinfo {author}
		{\bibfnamefont {L.}~\bibnamefont {Yu}}, \bibinfo {author} {\bibfnamefont
			{G.}~\bibnamefont {Liu}}, \bibinfo {author} {\bibfnamefont {X.}~\bibnamefont
			{Dong}}, \bibinfo {author} {\bibfnamefont {J.}~\bibnamefont {Zhang}},
		\bibinfo {author} {\bibfnamefont {C.}~\bibnamefont {Chen}}, \bibinfo {author}
		{\bibfnamefont {Z.}~\bibnamefont {Xu}}, \bibinfo {author} {\bibfnamefont
			{X.}~\bibnamefont {Chen}}, \bibinfo {author} {\bibfnamefont {X.}~\bibnamefont
			{Ma}}, \bibinfo {author} {\bibfnamefont {Q.}~\bibnamefont {Xue}},\ and\
		\bibinfo {author} {\bibfnamefont {X.~J.}\ \bibnamefont {Zhou}},\ }\bibfield
	{title} {\bibinfo {title} {Phase diagram and electronic indication of
			high-temperature superconductivity at {65 K} in single-layer {FeSe} films},\
	}\href {https://doi.org/10.1038/NMAT3648} {\bibfield  {journal} {\bibinfo
			{journal} {Nat. Mater.}\ }\textbf {\bibinfo {volume} {12}},\ \bibinfo {pages}
		{605} (\bibinfo {year} {2013})}\BibitemShut {NoStop}%
	\bibitem [{\citenamefont {Tana}\ \emph {et~al.}(2013)\citenamefont {Tana},
		\citenamefont {Zhang}, \citenamefont {Xia}, \citenamefont {Ye}, \citenamefont
		{Chen}, \citenamefont {Xie}, \citenamefont {Peng}, \citenamefont {Xu},
		\citenamefont {Fan}, \citenamefont {Xu}, \citenamefont {Jiang}, \citenamefont
		{Zhang}, \citenamefont {Lai}, \citenamefont {Xiang}, \citenamefont {Hu},
		\citenamefont {Xie},\ and\ \citenamefont {Feng}}]{FS_FeSe2}%
	\BibitemOpen
	\bibfield  {author} {\bibinfo {author} {\bibfnamefont {S.}~\bibnamefont
			{Tana}}, \bibinfo {author} {\bibfnamefont {Y.}~\bibnamefont {Zhang}},
		\bibinfo {author} {\bibfnamefont {M.}~\bibnamefont {Xia}}, \bibinfo {author}
		{\bibfnamefont {Z.}~\bibnamefont {Ye}}, \bibinfo {author} {\bibfnamefont
			{F.}~\bibnamefont {Chen}}, \bibinfo {author} {\bibfnamefont {X.}~\bibnamefont
			{Xie}}, \bibinfo {author} {\bibfnamefont {R.}~\bibnamefont {Peng}}, \bibinfo
		{author} {\bibfnamefont {D.}~\bibnamefont {Xu}}, \bibinfo {author}
		{\bibfnamefont {Q.}~\bibnamefont {Fan}}, \bibinfo {author} {\bibfnamefont
			{H.}~\bibnamefont {Xu}}, \bibinfo {author} {\bibfnamefont {J.}~\bibnamefont
			{Jiang}}, \bibinfo {author} {\bibfnamefont {T.}~\bibnamefont {Zhang}},
		\bibinfo {author} {\bibfnamefont {X.}~\bibnamefont {Lai}}, \bibinfo {author}
		{\bibfnamefont {T.}~\bibnamefont {Xiang}}, \bibinfo {author} {\bibfnamefont
			{J.}~\bibnamefont {Hu}}, \bibinfo {author} {\bibfnamefont {B.}~\bibnamefont
			{Xie}},\ and\ \bibinfo {author} {\bibfnamefont {D.}~\bibnamefont {Feng}},\
	}\bibfield  {title} {\bibinfo {title} {Interface-induced superconductivity
			and strain-dependent spin density waves in {FeSe/SrTiO$_3$} thin films},\
	}\href {https://doi.org/10.1038/NMAT3654} {\bibfield  {journal} {\bibinfo
			{journal} {Nat. Mater.}\ }\textbf {\bibinfo {volume} {12}},\ \bibinfo {pages}
		{634} (\bibinfo {year} {2013})}\BibitemShut {NoStop}%
	\bibitem [{\citenamefont {Si}\ \emph {et~al.}(2016)\citenamefont {Si},
		\citenamefont {Yu},\ and\ \citenamefont {Abrahams}}]{QMSi_Review}%
	\BibitemOpen
	\bibfield  {author} {\bibinfo {author} {\bibfnamefont {Q.~M.}\ \bibnamefont
			{Si}}, \bibinfo {author} {\bibfnamefont {R.}~\bibnamefont {Yu}},\ and\
		\bibinfo {author} {\bibfnamefont {E.}~\bibnamefont {Abrahams}},\ }\bibfield
	{title} {\bibinfo {title} {High-temperature superconductivity in iron
			pnictides and chalcogenides},\ }\href
	{https://doi.org/10.1038/natrevmats.2016.17} {\bibfield  {journal} {\bibinfo
			{journal} {Nat. Rev. Mater.}\ }\textbf {\bibinfo {volume} {1}},\ \bibinfo
		{pages} {16017} (\bibinfo {year} {2016})}\BibitemShut {NoStop}%
	\bibitem [{\citenamefont {Zhao}\ \emph {et~al.}(2016)\citenamefont {Zhao},
		\citenamefont {Liang}, \citenamefont {Yuan}, \citenamefont {Hu},
		\citenamefont {Liu}, \citenamefont {Huang}, \citenamefont {He}, \citenamefont
		{Shen}, \citenamefont {Xu}, \citenamefont {Liu}, \citenamefont {Yu},
		\citenamefont {Liu}, \citenamefont {Zhou}, \citenamefont {Huang},
		\citenamefont {Dong}, \citenamefont {Zhou}, \citenamefont {Liu},
		\citenamefont {Lu}, \citenamefont {Zhao}, \citenamefont {Chen}, \citenamefont
		{Xu},\ and\ \citenamefont {Zhou}}]{FS_LiFeO1}%
	\BibitemOpen
	\bibfield  {author} {\bibinfo {author} {\bibfnamefont {L.}~\bibnamefont
			{Zhao}}, \bibinfo {author} {\bibfnamefont {A.}~\bibnamefont {Liang}},
		\bibinfo {author} {\bibfnamefont {D.}~\bibnamefont {Yuan}}, \bibinfo {author}
		{\bibfnamefont {Y.}~\bibnamefont {Hu}}, \bibinfo {author} {\bibfnamefont
			{D.}~\bibnamefont {Liu}}, \bibinfo {author} {\bibfnamefont {J.}~\bibnamefont
			{Huang}}, \bibinfo {author} {\bibfnamefont {S.}~\bibnamefont {He}}, \bibinfo
		{author} {\bibfnamefont {B.}~\bibnamefont {Shen}}, \bibinfo {author}
		{\bibfnamefont {Y.}~\bibnamefont {Xu}}, \bibinfo {author} {\bibfnamefont
			{X.}~\bibnamefont {Liu}}, \bibinfo {author} {\bibfnamefont {L.}~\bibnamefont
			{Yu}}, \bibinfo {author} {\bibfnamefont {G.}~\bibnamefont {Liu}}, \bibinfo
		{author} {\bibfnamefont {H.}~\bibnamefont {Zhou}}, \bibinfo {author}
		{\bibfnamefont {Y.}~\bibnamefont {Huang}}, \bibinfo {author} {\bibfnamefont
			{X.}~\bibnamefont {Dong}}, \bibinfo {author} {\bibfnamefont {F.}~\bibnamefont
			{Zhou}}, \bibinfo {author} {\bibfnamefont {K.}~\bibnamefont {Liu}}, \bibinfo
		{author} {\bibfnamefont {Z.}~\bibnamefont {Lu}}, \bibinfo {author}
		{\bibfnamefont {Z.}~\bibnamefont {Zhao}}, \bibinfo {author} {\bibfnamefont
			{C.}~\bibnamefont {Chen}}, \bibinfo {author} {\bibfnamefont {Z.}~\bibnamefont
			{Xu}},\ and\ \bibinfo {author} {\bibfnamefont {X.~J.}\ \bibnamefont {Zhou}},\
	}\bibfield  {title} {\bibinfo {title} {Common electronic origin of
			superconductivity in {(Li,Fe)OHFeSe} bulk superconductor and single-layer
			{FeSe/SrTiO$_3$} films},\ }\href {https://doi.org/10.1038/ncomms10608}
	{\bibfield  {journal} {\bibinfo  {journal} {Nat. Commun.}\ }\textbf {\bibinfo
			{volume} {7}},\ \bibinfo {pages} {10608} (\bibinfo {year}
		{2016})}\BibitemShut {NoStop}%
	\bibitem [{\citenamefont {Niu}\ \emph {et~al.}(2015)\citenamefont {Niu},
		\citenamefont {Peng}, \citenamefont {Xu}, \citenamefont {Yan}, \citenamefont
		{Jiang}, \citenamefont {Xu}, \citenamefont {Yu}, \citenamefont {Song},
		\citenamefont {Huang}, \citenamefont {X.Wang}, \citenamefont {Xie},
		\citenamefont {Lu}, \citenamefont {Wang}, \citenamefont {Chen}, \citenamefont
		{Sun},\ and\ \citenamefont {Feng}}]{FS_LiFeO2}%
	\BibitemOpen
	\bibfield  {author} {\bibinfo {author} {\bibfnamefont {X.~H.}\ \bibnamefont
			{Niu}}, \bibinfo {author} {\bibfnamefont {R.}~\bibnamefont {Peng}}, \bibinfo
		{author} {\bibfnamefont {H.~C.}\ \bibnamefont {Xu}}, \bibinfo {author}
		{\bibfnamefont {Y.~J.}\ \bibnamefont {Yan}}, \bibinfo {author} {\bibfnamefont
			{J.}~\bibnamefont {Jiang}}, \bibinfo {author} {\bibfnamefont {D.~F.}\
			\bibnamefont {Xu}}, \bibinfo {author} {\bibfnamefont {T.~L.}\ \bibnamefont
			{Yu}}, \bibinfo {author} {\bibfnamefont {Q.}~\bibnamefont {Song}}, \bibinfo
		{author} {\bibfnamefont {Z.~C.}\ \bibnamefont {Huang}}, \bibinfo {author}
		{\bibfnamefont {Y.}~\bibnamefont {X.Wang}}, \bibinfo {author} {\bibfnamefont
			{B.~P.}\ \bibnamefont {Xie}}, \bibinfo {author} {\bibfnamefont {X.~F.}\
			\bibnamefont {Lu}}, \bibinfo {author} {\bibfnamefont {N.~Z.}\ \bibnamefont
			{Wang}}, \bibinfo {author} {\bibfnamefont {X.~H.}\ \bibnamefont {Chen}},
		\bibinfo {author} {\bibfnamefont {Z.}~\bibnamefont {Sun}},\ and\ \bibinfo
		{author} {\bibfnamefont {D.~L.}\ \bibnamefont {Feng}},\ }\bibfield  {title}
	{\bibinfo {title} {Surface electronic structure and isotropic superconducting
			gap in {(Li$_{0.8}$Fe$_{0.2}$)OHFeSe}},\ }\href
	{https://doi.org/10.1103/PhysRevB.92.060504} {\bibfield  {journal} {\bibinfo
			{journal} {Phys. Rev. B}\ }\textbf {\bibinfo {volume} {92}},\ \bibinfo
		{pages} {060504} (\bibinfo {year} {2015})}\BibitemShut {NoStop}%
	\bibitem [{\citenamefont {Du}\ \emph {et~al.}(2016)\citenamefont {Du},
		\citenamefont {Yang}, \citenamefont {Lin}, \citenamefont {Fang},
		\citenamefont {Du}, \citenamefont {Xing}, \citenamefont {Yang}, \citenamefont
		{Zhu},\ and\ \citenamefont {Wen}}]{FS_LiFeO3}%
	\BibitemOpen
	\bibfield  {author} {\bibinfo {author} {\bibfnamefont {Z.}~\bibnamefont
			{Du}}, \bibinfo {author} {\bibfnamefont {X.}~\bibnamefont {Yang}}, \bibinfo
		{author} {\bibfnamefont {H.}~\bibnamefont {Lin}}, \bibinfo {author}
		{\bibfnamefont {D.}~\bibnamefont {Fang}}, \bibinfo {author} {\bibfnamefont
			{G.}~\bibnamefont {Du}}, \bibinfo {author} {\bibfnamefont {J.}~\bibnamefont
			{Xing}}, \bibinfo {author} {\bibfnamefont {H.}~\bibnamefont {Yang}}, \bibinfo
		{author} {\bibfnamefont {X.}~\bibnamefont {Zhu}},\ and\ \bibinfo {author}
		{\bibfnamefont {H.-H.}\ \bibnamefont {Wen}},\ }\bibfield  {title} {\bibinfo
		{title} {Scrutinizing the double superconducting gaps and strong coupling
			pairing in {(Li$_{1-x}$Fe$_x$)OHFeSe}},\ }\href
	{https://doi.org/10.1038/ncomms10565} {\bibfield  {journal} {\bibinfo
			{journal} {Nat. Commun.}\ }\textbf {\bibinfo {volume} {7}},\ \bibinfo {pages}
		{10565} (\bibinfo {year} {2016})}\BibitemShut {NoStop}%
	\bibitem [{\citenamefont {Yan}\ \emph {et~al.}(2016)\citenamefont {Yan},
		\citenamefont {Zhang}, \citenamefont {Ren}, \citenamefont {Liu},
		\citenamefont {Lu}, \citenamefont {Wang}, \citenamefont {Niu}, \citenamefont
		{Fan}, \citenamefont {Miao}, \citenamefont {Tao}, \citenamefont {Xie},
		\citenamefont {Chen}, \citenamefont {Zhang},\ and\ \citenamefont
		{Feng}}]{FS_LiFeO4}%
	\BibitemOpen
	\bibfield  {author} {\bibinfo {author} {\bibfnamefont {Y.~J.}\ \bibnamefont
			{Yan}}, \bibinfo {author} {\bibfnamefont {W.~H.}\ \bibnamefont {Zhang}},
		\bibinfo {author} {\bibfnamefont {M.~Q.}\ \bibnamefont {Ren}}, \bibinfo
		{author} {\bibfnamefont {X.}~\bibnamefont {Liu}}, \bibinfo {author}
		{\bibfnamefont {X.~F.}\ \bibnamefont {Lu}}, \bibinfo {author} {\bibfnamefont
			{N.~Z.}\ \bibnamefont {Wang}}, \bibinfo {author} {\bibfnamefont {X.~H.}\
			\bibnamefont {Niu}}, \bibinfo {author} {\bibfnamefont {Q.}~\bibnamefont
			{Fan}}, \bibinfo {author} {\bibfnamefont {J.}~\bibnamefont {Miao}}, \bibinfo
		{author} {\bibfnamefont {R.}~\bibnamefont {Tao}}, \bibinfo {author}
		{\bibfnamefont {B.~P.}\ \bibnamefont {Xie}}, \bibinfo {author} {\bibfnamefont
			{X.~H.}\ \bibnamefont {Chen}}, \bibinfo {author} {\bibfnamefont
			{T.}~\bibnamefont {Zhang}},\ and\ \bibinfo {author} {\bibfnamefont {D.~L.}\
			\bibnamefont {Feng}},\ }\bibfield  {title} {\bibinfo {title} {Surface
			electronic structure and evidence of plain \textit{s}-wave superconductivity
			in {(Li$_{0.8}$Fe$_{0.2}$)OHFeSe}},\ }\href
	{https://doi.org/10.1103/PhysRevB.94.134502} {\bibfield  {journal} {\bibinfo
			{journal} {Phys. Rev. B}\ }\textbf {\bibinfo {volume} {94}},\ \bibinfo
		{pages} {134502} (\bibinfo {year} {2016})}\BibitemShut {NoStop}%
	\bibitem [{\citenamefont {Pan}\ \emph {et~al.}(2017)\citenamefont {Pan},
		\citenamefont {Shen}, \citenamefont {Hu}, \citenamefont {Feng}, \citenamefont
		{Park}, \citenamefont {Christianson}, \citenamefont {Wang}, \citenamefont
		{Hao}, \citenamefont {Wo}, \citenamefont {Yin}, \citenamefont {Maier},\ and\
		\citenamefont {Zhao}}]{PanNC}%
	\BibitemOpen
	\bibfield  {author} {\bibinfo {author} {\bibfnamefont {B.}~\bibnamefont
			{Pan}}, \bibinfo {author} {\bibfnamefont {Y.}~\bibnamefont {Shen}}, \bibinfo
		{author} {\bibfnamefont {D.}~\bibnamefont {Hu}}, \bibinfo {author}
		{\bibfnamefont {Y.}~\bibnamefont {Feng}}, \bibinfo {author} {\bibfnamefont
			{J.~T.}\ \bibnamefont {Park}}, \bibinfo {author} {\bibfnamefont {A.~D.}\
			\bibnamefont {Christianson}}, \bibinfo {author} {\bibfnamefont
			{Q.}~\bibnamefont {Wang}}, \bibinfo {author} {\bibfnamefont {Y.}~\bibnamefont
			{Hao}}, \bibinfo {author} {\bibfnamefont {H.}~\bibnamefont {Wo}}, \bibinfo
		{author} {\bibfnamefont {Z.}~\bibnamefont {Yin}}, \bibinfo {author}
		{\bibfnamefont {T.~A.}\ \bibnamefont {Maier}},\ and\ \bibinfo {author}
		{\bibfnamefont {J.}~\bibnamefont {Zhao}},\ }\bibfield  {title} {\bibinfo
		{title} {Structure of spin excitations in heavily electron-doped
			{Li$_{0.8}$Fe$_{0.2}$ODFeSe} superconductors},\ }\href
	{https://doi.org/10.1038/s41467-017-00162-x} {\bibfield  {journal} {\bibinfo
			{journal} {Nat. Commum.}\ }\textbf {\bibinfo {volume} {8}},\ \bibinfo {pages}
		{123} (\bibinfo {year} {2017})}\BibitemShut {NoStop}%
	\bibitem [{\citenamefont {Zhao}\ \emph {et~al.}(2009)\citenamefont {Zhao},
		\citenamefont {Adroja}, \citenamefont {Yao}, \citenamefont {Bewley},
		\citenamefont {Li}, \citenamefont {Wang}, \citenamefont {Wu}, \citenamefont
		{Chen}, \citenamefont {Hu},\ and\ \citenamefont {Dai}}]{CaFeAs_SW}%
	\BibitemOpen
	\bibfield  {author} {\bibinfo {author} {\bibfnamefont {J.}~\bibnamefont
			{Zhao}}, \bibinfo {author} {\bibfnamefont {D.~T.}\ \bibnamefont {Adroja}},
		\bibinfo {author} {\bibfnamefont {D.-X.}\ \bibnamefont {Yao}}, \bibinfo
		{author} {\bibfnamefont {R.}~\bibnamefont {Bewley}}, \bibinfo {author}
		{\bibfnamefont {S.}~\bibnamefont {Li}}, \bibinfo {author} {\bibfnamefont
			{X.~F.}\ \bibnamefont {Wang}}, \bibinfo {author} {\bibfnamefont
			{G.}~\bibnamefont {Wu}}, \bibinfo {author} {\bibfnamefont {X.~H.}\
			\bibnamefont {Chen}}, \bibinfo {author} {\bibfnamefont {J.}~\bibnamefont
			{Hu}},\ and\ \bibinfo {author} {\bibfnamefont {P.}~\bibnamefont {Dai}},\
	}\bibfield  {title} {\bibinfo {title} {Spin waves and magnetic exchange
			interactions in {CaFe$_2$As$_2$}},\ }\href
	{https://doi.org/10.1038/nphys1336} {\bibfield  {journal} {\bibinfo
			{journal} {Nat. Phys.}\ }\textbf {\bibinfo {volume} {5}},\ \bibinfo {pages}
		{555} (\bibinfo {year} {2009})}\BibitemShut {NoStop}%
	\bibitem [{\citenamefont {Liu}\ \emph {et~al.}(2012{\natexlab{a}})\citenamefont
		{Liu}, \citenamefont {Zhang}, \citenamefont {Mou}, \citenamefont {He},
		\citenamefont {Ou}, \citenamefont {Wang}, \citenamefont {Li}, \citenamefont
		{Wang}, \citenamefont {Zhao}, \citenamefont {He}, \citenamefont {Peng},
		\citenamefont {Liu}, \citenamefont {Chen}, \citenamefont {Yu}, \citenamefont
		{Liu}, \citenamefont {Dong}, \citenamefont {Zhang}, \citenamefont {Chen},
		\citenamefont {Xu}, \citenamefont {Hu}, \citenamefont {Chen}, \citenamefont
		{Ma}, \citenamefont {Xue},\ and\ \citenamefont {Zhou}}]{FeSe_film_Tc2}%
	\BibitemOpen
	\bibfield  {author} {\bibinfo {author} {\bibfnamefont {D.}~\bibnamefont
			{Liu}}, \bibinfo {author} {\bibfnamefont {W.}~\bibnamefont {Zhang}}, \bibinfo
		{author} {\bibfnamefont {D.}~\bibnamefont {Mou}}, \bibinfo {author}
		{\bibfnamefont {J.}~\bibnamefont {He}}, \bibinfo {author} {\bibfnamefont
			{Y.-B.}\ \bibnamefont {Ou}}, \bibinfo {author} {\bibfnamefont {Q.-Y.}\
			\bibnamefont {Wang}}, \bibinfo {author} {\bibfnamefont {Z.}~\bibnamefont
			{Li}}, \bibinfo {author} {\bibfnamefont {L.}~\bibnamefont {Wang}}, \bibinfo
		{author} {\bibfnamefont {L.}~\bibnamefont {Zhao}}, \bibinfo {author}
		{\bibfnamefont {S.}~\bibnamefont {He}}, \bibinfo {author} {\bibfnamefont
			{Y.}~\bibnamefont {Peng}}, \bibinfo {author} {\bibfnamefont {X.}~\bibnamefont
			{Liu}}, \bibinfo {author} {\bibfnamefont {C.}~\bibnamefont {Chen}}, \bibinfo
		{author} {\bibfnamefont {L.}~\bibnamefont {Yu}}, \bibinfo {author}
		{\bibfnamefont {G.}~\bibnamefont {Liu}}, \bibinfo {author} {\bibfnamefont
			{X.}~\bibnamefont {Dong}}, \bibinfo {author} {\bibfnamefont {J.}~\bibnamefont
			{Zhang}}, \bibinfo {author} {\bibfnamefont {C.}~\bibnamefont {Chen}},
		\bibinfo {author} {\bibfnamefont {Z.}~\bibnamefont {Xu}}, \bibinfo {author}
		{\bibfnamefont {J.}~\bibnamefont {Hu}}, \bibinfo {author} {\bibfnamefont
			{X.}~\bibnamefont {Chen}}, \bibinfo {author} {\bibfnamefont {X.}~\bibnamefont
			{Ma}}, \bibinfo {author} {\bibfnamefont {Q.}~\bibnamefont {Xue}},\ and\
		\bibinfo {author} {\bibfnamefont {X.~J.}\ \bibnamefont {Zhou}},\ }\bibfield
	{title} {\bibinfo {title} {Electronic origin of high-temperature
			superconductivity in single-layer {FeSe} superconductor},\ }\href
	{https://doi.org/10.1038/ncomms1946} {\bibfield  {journal} {\bibinfo
			{journal} {Nat. Commun.}\ }\textbf {\bibinfo {volume} {3}},\ \bibinfo {pages}
		{931} (\bibinfo {year} {2012}{\natexlab{a}})}\BibitemShut {NoStop}%
	\bibitem [{\citenamefont {Lei}\ \emph {et~al.}(2016)\citenamefont {Lei},
		\citenamefont {Xiang}, \citenamefont {Lu}, \citenamefont {Wang},
		\citenamefont {Chang}, \citenamefont {Shang}, \citenamefont {Zhang},
		\citenamefont {Zhang}, \citenamefont {Luo}, \citenamefont {Wu}, \citenamefont
		{Sun},\ and\ \citenamefont {Chen}}]{Gating}%
	\BibitemOpen
	\bibfield  {author} {\bibinfo {author} {\bibfnamefont {B.}~\bibnamefont
			{Lei}}, \bibinfo {author} {\bibfnamefont {Z.~J.}\ \bibnamefont {Xiang}},
		\bibinfo {author} {\bibfnamefont {X.~F.}\ \bibnamefont {Lu}}, \bibinfo
		{author} {\bibfnamefont {N.~Z.}\ \bibnamefont {Wang}}, \bibinfo {author}
		{\bibfnamefont {J.~R.}\ \bibnamefont {Chang}}, \bibinfo {author}
		{\bibfnamefont {C.}~\bibnamefont {Shang}}, \bibinfo {author} {\bibfnamefont
			{A.~M.}\ \bibnamefont {Zhang}}, \bibinfo {author} {\bibfnamefont {Q.~M.}\
			\bibnamefont {Zhang}}, \bibinfo {author} {\bibfnamefont {X.~G.}\ \bibnamefont
			{Luo}}, \bibinfo {author} {\bibfnamefont {T.}~\bibnamefont {Wu}}, \bibinfo
		{author} {\bibfnamefont {Z.}~\bibnamefont {Sun}},\ and\ \bibinfo {author}
		{\bibfnamefont {X.~H.}\ \bibnamefont {Chen}},\ }\bibfield  {title} {\bibinfo
		{title} {Gate-tuned superconductor-insulator transition in {(Li,Fe)OHFeSe}},\
	}\href {https://doi.org/10.1103/PhysRevB.93.060501} {\bibfield  {journal}
		{\bibinfo  {journal} {Phys. Rev. B}\ }\textbf {\bibinfo {volume} {93}},\
		\bibinfo {pages} {060501} (\bibinfo {year} {2016})}\BibitemShut {NoStop}%
	\bibitem [{\citenamefont {Dong}\ \emph
		{et~al.}(2015{\natexlab{a}})\citenamefont {Dong}, \citenamefont {Zhou},
		\citenamefont {Yang}, \citenamefont {Yuan}, \citenamefont {Jin},
		\citenamefont {Zhou}, \citenamefont {Yuan}, \citenamefont {Wei},
		\citenamefont {Li}, \citenamefont {Wang}, \citenamefont {Zhang},\ and\
		\citenamefont {Zhao}}]{LiFeOIns1}%
	\BibitemOpen
	\bibfield  {author} {\bibinfo {author} {\bibfnamefont {X.}~\bibnamefont
			{Dong}}, \bibinfo {author} {\bibfnamefont {H.}~\bibnamefont {Zhou}}, \bibinfo
		{author} {\bibfnamefont {H.}~\bibnamefont {Yang}}, \bibinfo {author}
		{\bibfnamefont {J.}~\bibnamefont {Yuan}}, \bibinfo {author} {\bibfnamefont
			{K.}~\bibnamefont {Jin}}, \bibinfo {author} {\bibfnamefont {F.}~\bibnamefont
			{Zhou}}, \bibinfo {author} {\bibfnamefont {D.}~\bibnamefont {Yuan}}, \bibinfo
		{author} {\bibfnamefont {L.}~\bibnamefont {Wei}}, \bibinfo {author}
		{\bibfnamefont {J.}~\bibnamefont {Li}}, \bibinfo {author} {\bibfnamefont
			{X.}~\bibnamefont {Wang}}, \bibinfo {author} {\bibfnamefont {G.}~\bibnamefont
			{Zhang}},\ and\ \bibinfo {author} {\bibfnamefont {Z.}~\bibnamefont {Zhao}},\
	}\bibfield  {title} {\bibinfo {title} {Phase diagram of
			{(Li$_{1-x}$Fe$_x$)OHFeSe}: A bridge between iron selenide and arsenide
			superconductors},\ }\href {https://doi.org/10.1021/ja511292f} {\bibfield
		{journal} {\bibinfo  {journal} {J. Am. Chem. C}\ }\textbf {\bibinfo {volume}
			{137}},\ \bibinfo {pages} {66} (\bibinfo {year}
		{2015}{\natexlab{a}})}\BibitemShut {NoStop}%
	\bibitem [{\citenamefont {Sun}\ \emph {et~al.}(2015)\citenamefont {Sun},
		\citenamefont {Woodruff}, \citenamefont {Cassidy}, \citenamefont {Allcroft},
		\citenamefont {Sedlmaier}, \citenamefont {Thompson}, \citenamefont {Bingham},
		\citenamefont {Forder}, \citenamefont {Cartenet}, \citenamefont {Mary},
		\citenamefont {Ramos}, \citenamefont {Foronda}, \citenamefont {Williams},
		\citenamefont {Li}, \citenamefont {Blundell},\ and\ \citenamefont
		{Clarke}}]{LiFeOIns2}%
	\BibitemOpen
	\bibfield  {author} {\bibinfo {author} {\bibfnamefont {H.}~\bibnamefont
			{Sun}}, \bibinfo {author} {\bibfnamefont {D.~N.}\ \bibnamefont {Woodruff}},
		\bibinfo {author} {\bibfnamefont {S.~J.}\ \bibnamefont {Cassidy}}, \bibinfo
		{author} {\bibfnamefont {G.~M.}\ \bibnamefont {Allcroft}}, \bibinfo {author}
		{\bibfnamefont {S.~J.}\ \bibnamefont {Sedlmaier}}, \bibinfo {author}
		{\bibfnamefont {A.~L.}\ \bibnamefont {Thompson}}, \bibinfo {author}
		{\bibfnamefont {P.~A.}\ \bibnamefont {Bingham}}, \bibinfo {author}
		{\bibfnamefont {S.~D.}\ \bibnamefont {Forder}}, \bibinfo {author}
		{\bibfnamefont {S.}~\bibnamefont {Cartenet}}, \bibinfo {author}
		{\bibfnamefont {N.}~\bibnamefont {Mary}}, \bibinfo {author} {\bibfnamefont
			{S.}~\bibnamefont {Ramos}}, \bibinfo {author} {\bibfnamefont {F.~R.}\
			\bibnamefont {Foronda}}, \bibinfo {author} {\bibfnamefont {B.~H.}\
			\bibnamefont {Williams}}, \bibinfo {author} {\bibfnamefont {X.}~\bibnamefont
			{Li}}, \bibinfo {author} {\bibfnamefont {S.~J.}\ \bibnamefont {Blundell}},\
		and\ \bibinfo {author} {\bibfnamefont {S.~J.}\ \bibnamefont {Clarke}},\
	}\bibfield  {title} {\bibinfo {title} {Soft chemical control of
			superconductivity in lithium iron selenide hydroxides
			{Li$_{1-x}$Fe$_x$(OH)Fe$_{1-y}$Se}},\ }\href
	{https://doi.org/10.1021/ic5028702} {\bibfield  {journal} {\bibinfo
			{journal} {Inorg. Chem.}\ }\textbf {\bibinfo {volume} {54}},\ \bibinfo
		{pages} {1958} (\bibinfo {year} {2015})}\BibitemShut {NoStop}%
	\bibitem [{\citenamefont {Hu}\ \emph {et~al.}(2019)\citenamefont {Hu},
		\citenamefont {Wang}, \citenamefont {Shi}, \citenamefont {Meng},
		\citenamefont {Shang}, \citenamefont {Ma}, \citenamefont {Luo},\ and\
		\citenamefont {Chen}}]{LiFeOIns3}%
	\BibitemOpen
	\bibfield  {author} {\bibinfo {author} {\bibfnamefont {G.~B.}\ \bibnamefont
			{Hu}}, \bibinfo {author} {\bibfnamefont {N.~Z.}\ \bibnamefont {Wang}},
		\bibinfo {author} {\bibfnamefont {M.~Z.}\ \bibnamefont {Shi}}, \bibinfo
		{author} {\bibfnamefont {F.~B.}\ \bibnamefont {Meng}}, \bibinfo {author}
		{\bibfnamefont {C.}~\bibnamefont {Shang}}, \bibinfo {author} {\bibfnamefont
			{L.~K.}\ \bibnamefont {Ma}}, \bibinfo {author} {\bibfnamefont {X.~G.}\
			\bibnamefont {Luo}},\ and\ \bibinfo {author} {\bibfnamefont {X.~H.}\
			\bibnamefont {Chen}},\ }\bibfield  {title} {\bibinfo {title}
		{Superconductivity in solid-state synthesized {(Li,Fe)OHFeSe} by tuning {Fe}
			vacancies in {FeSe} layer},\ }\href
	{https://doi.org/10.1103/PhysRevMaterials.3.064802} {\bibfield  {journal}
		{\bibinfo  {journal} {Phys. Rev. Materials}\ }\textbf {\bibinfo {volume}
			{3}},\ \bibinfo {pages} {064802} (\bibinfo {year} {2019})}\BibitemShut
	{NoStop}%
	\bibitem [{\citenamefont {Zhou}\ \emph {et~al.}(2016)\citenamefont {Zhou},
		\citenamefont {Borg}, \citenamefont {Lynn}, \citenamefont {Saha},
		\citenamefont {Paglione},\ and\ \citenamefont {Rodriguez}}]{LiFeONPD}%
	\BibitemOpen
	\bibfield  {author} {\bibinfo {author} {\bibfnamefont {X.~Q.}\ \bibnamefont
			{Zhou}}, \bibinfo {author} {\bibfnamefont {C.~K.~H.}\ \bibnamefont {Borg}},
		\bibinfo {author} {\bibfnamefont {J.~W.}\ \bibnamefont {Lynn}}, \bibinfo
		{author} {\bibfnamefont {S.~R.}\ \bibnamefont {Saha}}, \bibinfo {author}
		{\bibfnamefont {J.}~\bibnamefont {Paglione}},\ and\ \bibinfo {author}
		{\bibfnamefont {E.~E.}\ \bibnamefont {Rodriguez}},\ }\bibfield  {title}
	{\bibinfo {title} {The preparation and phase diagrams of
			{($^7$Li$_{1-x}$Fe$_x$OD)FeSe} and {(Li$_{1-x}$Fe$_x$OH)FeSe}
			superconductors},\ }\href {https://doi.org/10.1039/C5TC04041H} {\bibfield
		{journal} {\bibinfo  {journal} {J. Mater. Chem. C}\ }\textbf {\bibinfo
			{volume} {4}},\ \bibinfo {pages} {3934} (\bibinfo {year} {2016})}\BibitemShut
	{NoStop}%
	\bibitem [{Sup()}]{Supplement}%
	\BibitemOpen
	\href@noop {} {}\bibinfo {note} {See Supplemental Material at
		[URL-will-be-inserted-by-publisher] for detailed information on sample
		characterizations, experimental configuration and background subtraction,
		which includes Refs. [2,6,7,17,19,28-30,32,33].}\BibitemShut {Stop}%
	\bibitem [{\citenamefont {Dong}\ \emph
		{et~al.}(2015{\natexlab{b}})\citenamefont {Dong}, \citenamefont {Jin},
		\citenamefont {Yuan}, \citenamefont {Zhou}, \citenamefont {Yuan},
		\citenamefont {Huang}, \citenamefont {Hua}, \citenamefont {Sun},
		\citenamefont {Zheng}, \citenamefont {Hu}, \citenamefont {Mao}, \citenamefont
		{Ma}, \citenamefont {Zhang}, \citenamefont {Zhou},\ and\ \citenamefont
		{Zhao}}]{XiaoliDong}%
	\BibitemOpen
	\bibfield  {author} {\bibinfo {author} {\bibfnamefont {X.}~\bibnamefont
			{Dong}}, \bibinfo {author} {\bibfnamefont {K.}~\bibnamefont {Jin}}, \bibinfo
		{author} {\bibfnamefont {D.}~\bibnamefont {Yuan}}, \bibinfo {author}
		{\bibfnamefont {H.}~\bibnamefont {Zhou}}, \bibinfo {author} {\bibfnamefont
			{J.}~\bibnamefont {Yuan}}, \bibinfo {author} {\bibfnamefont {Y.}~\bibnamefont
			{Huang}}, \bibinfo {author} {\bibfnamefont {W.}~\bibnamefont {Hua}}, \bibinfo
		{author} {\bibfnamefont {J.}~\bibnamefont {Sun}}, \bibinfo {author}
		{\bibfnamefont {P.}~\bibnamefont {Zheng}}, \bibinfo {author} {\bibfnamefont
			{W.}~\bibnamefont {Hu}}, \bibinfo {author} {\bibfnamefont {Y.}~\bibnamefont
			{Mao}}, \bibinfo {author} {\bibfnamefont {M.}~\bibnamefont {Ma}}, \bibinfo
		{author} {\bibfnamefont {G.}~\bibnamefont {Zhang}}, \bibinfo {author}
		{\bibfnamefont {F.}~\bibnamefont {Zhou}},\ and\ \bibinfo {author}
		{\bibfnamefont {Z.}~\bibnamefont {Zhao}},\ }\bibfield  {title} {\bibinfo
		{title} {{(Li$_{0.84}$Fe$_{0.16}$)OHFe$_{0.98}$Se} superconductor:
			{Ion}-exchange synthesis of large single-crystal and highly two-dimensional
			electron properties},\ }\href {https://doi.org/10.1103/PhysRevB.92.064515}
	{\bibfield  {journal} {\bibinfo  {journal} {Phys. Rev. B}\ }\textbf {\bibinfo
			{volume} {92}},\ \bibinfo {pages} {064515} (\bibinfo {year}
		{2015}{\natexlab{b}})}\BibitemShut {NoStop}%
	\bibitem [{\citenamefont {Hu}\ \emph {et~al.}(2021)\citenamefont {Hu},
		\citenamefont {Feng}, \citenamefont {Park}, \citenamefont {Wo}, \citenamefont
		{Wang}, \citenamefont {Bourdarot}, \citenamefont {Ivanov},\ and\
		\citenamefont {Zhao}}]{Polarize}%
	\BibitemOpen
	\bibfield  {author} {\bibinfo {author} {\bibfnamefont {D.}~\bibnamefont
			{Hu}}, \bibinfo {author} {\bibfnamefont {Y.}~\bibnamefont {Feng}}, \bibinfo
		{author} {\bibfnamefont {J.~T.}\ \bibnamefont {Park}}, \bibinfo {author}
		{\bibfnamefont {H.}~\bibnamefont {Wo}}, \bibinfo {author} {\bibfnamefont
			{Q.}~\bibnamefont {Wang}}, \bibinfo {author} {\bibfnamefont {F.}~\bibnamefont
			{Bourdarot}}, \bibinfo {author} {\bibfnamefont {A.}~\bibnamefont {Ivanov}},\
		and\ \bibinfo {author} {\bibfnamefont {J.}~\bibnamefont {Zhao}},\ }\bibfield
	{title} {\bibinfo {title} {Polarized neutron scattering studies of magnetic
			excitations in iron-selenide superconductor {Li$_{0.8}$Fe$_{0.2}$ODFeSe}
			({{\textit{T}$_c$} = 41 K})},\ }\href
	{https://doi.org/10.1088/1361-648X/ac1d16} {\bibfield  {journal} {\bibinfo
			{journal} {J. Phys.: Condens. Matter}\ }\textbf {\bibinfo {volume} {33}},\
		\bibinfo {pages} {45LT01} (\bibinfo {year} {2021})}\BibitemShut {NoStop}%
	\bibitem [{\citenamefont {Wo}\ \emph {et~al.}(2019)\citenamefont {Wo},
		\citenamefont {Wang}, \citenamefont {Shen}, \citenamefont {Zhang},
		\citenamefont {Hao}, \citenamefont {Feng}, \citenamefont {Shen},
		\citenamefont {He}, \citenamefont {Pan}, \citenamefont {Wang}, \citenamefont
		{Nakajima}, \citenamefont {Ohira-Kawamura}, \citenamefont {Steffens},
		\citenamefont {Boehm}, \citenamefont {Schmalzl}, \citenamefont {Forrest},
		\citenamefont {Matsuda}, \citenamefont {Zhao}, \citenamefont {Lynn},
		\citenamefont {Yin},\ and\ \citenamefont {Zhao}}]{YFG}%
	\BibitemOpen
	\bibfield  {author} {\bibinfo {author} {\bibfnamefont {H.}~\bibnamefont
			{Wo}}, \bibinfo {author} {\bibfnamefont {Q.}~\bibnamefont {Wang}}, \bibinfo
		{author} {\bibfnamefont {Y.}~\bibnamefont {Shen}}, \bibinfo {author}
		{\bibfnamefont {X.}~\bibnamefont {Zhang}}, \bibinfo {author} {\bibfnamefont
			{Y.}~\bibnamefont {Hao}}, \bibinfo {author} {\bibfnamefont {Y.}~\bibnamefont
			{Feng}}, \bibinfo {author} {\bibfnamefont {S.}~\bibnamefont {Shen}}, \bibinfo
		{author} {\bibfnamefont {Z.}~\bibnamefont {He}}, \bibinfo {author}
		{\bibfnamefont {B.}~\bibnamefont {Pan}}, \bibinfo {author} {\bibfnamefont
			{W.}~\bibnamefont {Wang}}, \bibinfo {author} {\bibfnamefont {K.}~\bibnamefont
			{Nakajima}}, \bibinfo {author} {\bibfnamefont {S.}~\bibnamefont
			{Ohira-Kawamura}}, \bibinfo {author} {\bibfnamefont {P.}~\bibnamefont
			{Steffens}}, \bibinfo {author} {\bibfnamefont {M.}~\bibnamefont {Boehm}},
		\bibinfo {author} {\bibfnamefont {K.}~\bibnamefont {Schmalzl}}, \bibinfo
		{author} {\bibfnamefont {T.~R.}\ \bibnamefont {Forrest}}, \bibinfo {author}
		{\bibfnamefont {M.}~\bibnamefont {Matsuda}}, \bibinfo {author} {\bibfnamefont
			{Y.}~\bibnamefont {Zhao}}, \bibinfo {author} {\bibfnamefont {J.~W.}\
			\bibnamefont {Lynn}}, \bibinfo {author} {\bibfnamefont {Z.}~\bibnamefont
			{Yin}},\ and\ \bibinfo {author} {\bibfnamefont {J.}~\bibnamefont {Zhao}},\
	}\bibfield  {title} {\bibinfo {title} {Coexistence of ferromagnetic and
			stripe-type antiferromagnetic spin fluctuations in {YFe$_2$Ge$_2$}},\ }\href
	{https://doi.org/10.1103/PhysRevLett.122.217003} {\bibfield  {journal}
		{\bibinfo  {journal} {Phys. Rev. Lett.}\ }\textbf {\bibinfo {volume} {122}},\
		\bibinfo {pages} {217003} (\bibinfo {year} {2019})}\BibitemShut {NoStop}%
	\bibitem [{\citenamefont {Davies}\ \emph {et~al.}(2016)\citenamefont {Davies},
		\citenamefont {Rahn}, \citenamefont {Walker}, \citenamefont {Ewings},
		\citenamefont {Woodruff}, \citenamefont {Clarke},\ and\ \citenamefont
		{Boothroyd}}]{Boothroyd}%
	\BibitemOpen
	\bibfield  {author} {\bibinfo {author} {\bibfnamefont {N.~R.}\ \bibnamefont
			{Davies}}, \bibinfo {author} {\bibfnamefont {M.~C.}\ \bibnamefont {Rahn}},
		\bibinfo {author} {\bibfnamefont {H.~C.}\ \bibnamefont {Walker}}, \bibinfo
		{author} {\bibfnamefont {R.~A.}\ \bibnamefont {Ewings}}, \bibinfo {author}
		{\bibfnamefont {D.~N.}\ \bibnamefont {Woodruff}}, \bibinfo {author}
		{\bibfnamefont {S.~J.}\ \bibnamefont {Clarke}},\ and\ \bibinfo {author}
		{\bibfnamefont {A.~T.}\ \bibnamefont {Boothroyd}},\ }\bibfield  {title}
	{\bibinfo {title} {Spin resonance in the superconducting state of
			{Li$_{1-\textit{x}}$Fe$_\textit{x}$ODFe$_{1-\textit{y}}$Se} observed by
			neutron spectroscopy},\ }\href {https://doi.org/10.1103/PhysRevB.94.144503}
	{\bibfield  {journal} {\bibinfo  {journal} {Phys. Rev. B}\ }\textbf {\bibinfo
			{volume} {94}},\ \bibinfo {pages} {144503} (\bibinfo {year}
		{2016})}\BibitemShut {NoStop}%
	\bibitem [{\citenamefont {Wang}\ \emph {et~al.}(2016)\citenamefont {Wang},
		\citenamefont {Shen}, \citenamefont {Pan}, \citenamefont {Zhang},
		\citenamefont {Ikeuchi}, \citenamefont {Iida}, \citenamefont {Christianson},
		\citenamefont {Walker}, \citenamefont {Adroja}, \citenamefont {Abdel-Hafiez},
		\citenamefont {Chen}, \citenamefont {Chareev}, \citenamefont {Vasiliev},\
		and\ \citenamefont {Zhao}}]{WangNC}%
	\BibitemOpen
	\bibfield  {author} {\bibinfo {author} {\bibfnamefont {Q.}~\bibnamefont
			{Wang}}, \bibinfo {author} {\bibfnamefont {Y.}~\bibnamefont {Shen}}, \bibinfo
		{author} {\bibfnamefont {B.}~\bibnamefont {Pan}}, \bibinfo {author}
		{\bibfnamefont {X.}~\bibnamefont {Zhang}}, \bibinfo {author} {\bibfnamefont
			{K.}~\bibnamefont {Ikeuchi}}, \bibinfo {author} {\bibfnamefont
			{K.}~\bibnamefont {Iida}}, \bibinfo {author} {\bibfnamefont {A.~D.}\
			\bibnamefont {Christianson}}, \bibinfo {author} {\bibfnamefont {H.~C.}\
			\bibnamefont {Walker}}, \bibinfo {author} {\bibfnamefont {D.~T.}\
			\bibnamefont {Adroja}}, \bibinfo {author} {\bibfnamefont {M.}~\bibnamefont
			{Abdel-Hafiez}}, \bibinfo {author} {\bibfnamefont {X.}~\bibnamefont {Chen}},
		\bibinfo {author} {\bibfnamefont {D.~A.}\ \bibnamefont {Chareev}}, \bibinfo
		{author} {\bibfnamefont {A.~N.}\ \bibnamefont {Vasiliev}},\ and\ \bibinfo
		{author} {\bibfnamefont {J.}~\bibnamefont {Zhao}},\ }\bibfield  {title}
	{\bibinfo {title} {Magnetic ground state of {FeSe}},\ }\href
	{https://doi.org/10.1038/ncomms12182} {\bibfield  {journal} {\bibinfo
			{journal} {Nat. Commum.}\ }\textbf {\bibinfo {volume} {7}},\ \bibinfo {pages}
		{12182} (\bibinfo {year} {2016})}\BibitemShut {NoStop}%
	\bibitem [{\citenamefont {Tam}\ \emph {et~al.}(2020)\citenamefont {Tam},
		\citenamefont {Yin}, \citenamefont {Xie}, \citenamefont {Wang}, \citenamefont
		{Stone}, \citenamefont {Adroja}, \citenamefont {Walker}, \citenamefont {Yi},\
		and\ \citenamefont {Dai}}]{TamPRB}%
	\BibitemOpen
	\bibfield  {author} {\bibinfo {author} {\bibfnamefont {D.~W.}\ \bibnamefont
			{Tam}}, \bibinfo {author} {\bibfnamefont {Z.}~\bibnamefont {Yin}}, \bibinfo
		{author} {\bibfnamefont {Y.}~\bibnamefont {Xie}}, \bibinfo {author}
		{\bibfnamefont {W.}~\bibnamefont {Wang}}, \bibinfo {author} {\bibfnamefont
			{M.~B.}\ \bibnamefont {Stone}}, \bibinfo {author} {\bibfnamefont {D.~T.}\
			\bibnamefont {Adroja}}, \bibinfo {author} {\bibfnamefont {H.~C.}\
			\bibnamefont {Walker}}, \bibinfo {author} {\bibfnamefont {M.}~\bibnamefont
			{Yi}},\ and\ \bibinfo {author} {\bibfnamefont {P.}~\bibnamefont {Dai}},\
	}\bibfield  {title} {\bibinfo {title} {Orbital selective spin waves in
			detwinned {NaFeAs}},\ }\href {https://doi.org/10.1103/PhysRevB.102.054430}
	{\bibfield  {journal} {\bibinfo  {journal} {Phys. Rev. B}\ }\textbf {\bibinfo
			{volume} {102}},\ \bibinfo {pages} {054430} (\bibinfo {year}
		{2020})}\BibitemShut {NoStop}%
	\bibitem [{\citenamefont {Harriger}\ \emph {et~al.}(2011)\citenamefont
		{Harriger}, \citenamefont {Luo}, \citenamefont {Liu}, \citenamefont {Frost},
		\citenamefont {Hu}, \citenamefont {Norman},\ and\ \citenamefont
		{Dai}}]{BaFeAs_SW}%
	\BibitemOpen
	\bibfield  {author} {\bibinfo {author} {\bibfnamefont {L.~W.}\ \bibnamefont
			{Harriger}}, \bibinfo {author} {\bibfnamefont {H.~Q.}\ \bibnamefont {Luo}},
		\bibinfo {author} {\bibfnamefont {M.~S.}\ \bibnamefont {Liu}}, \bibinfo
		{author} {\bibfnamefont {C.}~\bibnamefont {Frost}}, \bibinfo {author}
		{\bibfnamefont {J.~P.}\ \bibnamefont {Hu}}, \bibinfo {author} {\bibfnamefont
			{M.~R.}\ \bibnamefont {Norman}},\ and\ \bibinfo {author} {\bibfnamefont
			{P.}~\bibnamefont {Dai}},\ }\bibfield  {title} {\bibinfo {title} {Nematic
			spin fluid in the tetragonal phase of {BaFe$_2$As$_2$}},\ }\href
	{https://doi.org/10.1103/PhysRevB.84.054544} {\bibfield  {journal} {\bibinfo
			{journal} {Phys. Rev. B}\ }\textbf {\bibinfo {volume} {84}},\ \bibinfo
		{pages} {054544} (\bibinfo {year} {2011})}\BibitemShut {NoStop}%
	\bibitem [{\citenamefont {Zhang}\ \emph {et~al.}(2014)\citenamefont {Zhang},
		\citenamefont {Harriger}, \citenamefont {Yin}, \citenamefont {Lv},
		\citenamefont {Wang}, \citenamefont {Tan}, \citenamefont {Song},
		\citenamefont {Abernathy}, \citenamefont {Tian}, \citenamefont {Egami},
		\citenamefont {Haule}, \citenamefont {Kotliar},\ and\ \citenamefont
		{Dai}}]{NaFeAs_SW}%
	\BibitemOpen
	\bibfield  {author} {\bibinfo {author} {\bibfnamefont {C.}~\bibnamefont
			{Zhang}}, \bibinfo {author} {\bibfnamefont {L.~W.}\ \bibnamefont {Harriger}},
		\bibinfo {author} {\bibfnamefont {Z.}~\bibnamefont {Yin}}, \bibinfo {author}
		{\bibfnamefont {W.}~\bibnamefont {Lv}}, \bibinfo {author} {\bibfnamefont
			{M.}~\bibnamefont {Wang}}, \bibinfo {author} {\bibfnamefont {G.}~\bibnamefont
			{Tan}}, \bibinfo {author} {\bibfnamefont {Y.}~\bibnamefont {Song}}, \bibinfo
		{author} {\bibfnamefont {D.~L.}\ \bibnamefont {Abernathy}}, \bibinfo {author}
		{\bibfnamefont {W.}~\bibnamefont {Tian}}, \bibinfo {author} {\bibfnamefont
			{T.}~\bibnamefont {Egami}}, \bibinfo {author} {\bibfnamefont
			{K.}~\bibnamefont {Haule}}, \bibinfo {author} {\bibfnamefont
			{G.}~\bibnamefont {Kotliar}},\ and\ \bibinfo {author} {\bibfnamefont
			{P.}~\bibnamefont {Dai}},\ }\bibfield  {title} {\bibinfo {title} {Effect of
			pnictogen height on spin waves in iron pnictides},\ }\href
	{https://doi.org/10.1103/PhysRevLett.112.217202} {\bibfield  {journal}
		{\bibinfo  {journal} {Phys. Rev. Lett.}\ }\textbf {\bibinfo {volume} {112}},\
		\bibinfo {pages} {217202} (\bibinfo {year} {2014})}\BibitemShut {NoStop}%
	\bibitem [{\citenamefont {Hu}\ \emph {et~al.}(2016)\citenamefont {Hu},
		\citenamefont {Yin}, \citenamefont {Zhang}, \citenamefont {Ewings},
		\citenamefont {Ikeuchi}, \citenamefont {Nakamura}, \citenamefont {Roessli},
		\citenamefont {Wei}, \citenamefont {Zhao}, \citenamefont {Chen},
		\citenamefont {Li}, \citenamefont {Luo}, \citenamefont {Haule}, \citenamefont
		{Kotliar},\ and\ \citenamefont {Dai}}]{BaFeAsP}%
	\BibitemOpen
	\bibfield  {author} {\bibinfo {author} {\bibfnamefont {D.}~\bibnamefont
			{Hu}}, \bibinfo {author} {\bibfnamefont {Z.}~\bibnamefont {Yin}}, \bibinfo
		{author} {\bibfnamefont {W.}~\bibnamefont {Zhang}}, \bibinfo {author}
		{\bibfnamefont {R.~A.}\ \bibnamefont {Ewings}}, \bibinfo {author}
		{\bibfnamefont {K.}~\bibnamefont {Ikeuchi}}, \bibinfo {author} {\bibfnamefont
			{M.}~\bibnamefont {Nakamura}}, \bibinfo {author} {\bibfnamefont
			{B.}~\bibnamefont {Roessli}}, \bibinfo {author} {\bibfnamefont
			{Y.}~\bibnamefont {Wei}}, \bibinfo {author} {\bibfnamefont {L.}~\bibnamefont
			{Zhao}}, \bibinfo {author} {\bibfnamefont {G.}~\bibnamefont {Chen}}, \bibinfo
		{author} {\bibfnamefont {S.}~\bibnamefont {Li}}, \bibinfo {author}
		{\bibfnamefont {H.}~\bibnamefont {Luo}}, \bibinfo {author} {\bibfnamefont
			{K.}~\bibnamefont {Haule}}, \bibinfo {author} {\bibfnamefont
			{G.}~\bibnamefont {Kotliar}},\ and\ \bibinfo {author} {\bibfnamefont
			{P.}~\bibnamefont {Dai}},\ }\bibfield  {title} {\bibinfo {title} {Spin
			excitations in optimally {P-doped} {BaFe$_{2}$(As$_{0.7}$P$_{0.3}$)$_2$}
			superconductor},\ }\href {https://doi.org/10.1103/PhysRevB.94.094504}
	{\bibfield  {journal} {\bibinfo  {journal} {Phys. Rev. B}\ }\textbf {\bibinfo
			{volume} {94}},\ \bibinfo {pages} {094504} (\bibinfo {year}
		{2016})}\BibitemShut {NoStop}%
	\bibitem [{\citenamefont {Yin}\ \emph {et~al.}(2011)\citenamefont {Yin},
		\citenamefont {Haule},\ and\ \citenamefont {Kotliar}}]{Zhiping1}%
	\BibitemOpen
	\bibfield  {author} {\bibinfo {author} {\bibfnamefont {Z.~P.}\ \bibnamefont
			{Yin}}, \bibinfo {author} {\bibfnamefont {K.}~\bibnamefont {Haule}},\ and\
		\bibinfo {author} {\bibfnamefont {G.}~\bibnamefont {Kotliar}},\ }\bibfield
	{title} {\bibinfo {title} {Kinetic frustration and the nature of the magnetic
			and paramagnetic states in iron pnictides and iron chalcogenides},\ }\href
	{https://doi.org/10.1038/NMAT3120} {\bibfield  {journal} {\bibinfo  {journal}
			{Nat. Mater.}\ }\textbf {\bibinfo {volume} {10}},\ \bibinfo {pages} {932}
		(\bibinfo {year} {2011})}\BibitemShut {NoStop}%
	\bibitem [{\citenamefont {Yin}\ \emph {et~al.}(2014)\citenamefont {Yin},
		\citenamefont {Haule},\ and\ \citenamefont {Kotliar}}]{Zhiping2}%
	\BibitemOpen
	\bibfield  {author} {\bibinfo {author} {\bibfnamefont {Z.~P.}\ \bibnamefont
			{Yin}}, \bibinfo {author} {\bibfnamefont {K.}~\bibnamefont {Haule}},\ and\
		\bibinfo {author} {\bibfnamefont {G.}~\bibnamefont {Kotliar}},\ }\bibfield
	{title} {\bibinfo {title} {Spin dynamics and orbital-antiphase pairing
			symmetry in iron-based superconductors},\ }\href
	{https://doi.org/10.1038/NPHYS3116} {\bibfield  {journal} {\bibinfo
			{journal} {Nat. Phys.}\ }\textbf {\bibinfo {volume} {10}},\ \bibinfo {pages}
		{845} (\bibinfo {year} {2014})}\BibitemShut {NoStop}%
	\bibitem [{\citenamefont {Liu}\ \emph {et~al.}(2012{\natexlab{b}})\citenamefont
		{Liu}, \citenamefont {Harriger}, \citenamefont {Luo}, \citenamefont {Wang},
		\citenamefont {Ewings}, \citenamefont {Guidi}, \citenamefont {Park},
		\citenamefont {Haule}, \citenamefont {Kotliar}, \citenamefont {Hayden},\ and\
		\citenamefont {Dai}}]{122_moment}%
	\BibitemOpen
	\bibfield  {author} {\bibinfo {author} {\bibfnamefont {M.}~\bibnamefont
			{Liu}}, \bibinfo {author} {\bibfnamefont {L.~W.}\ \bibnamefont {Harriger}},
		\bibinfo {author} {\bibfnamefont {H.}~\bibnamefont {Luo}}, \bibinfo {author}
		{\bibfnamefont {M.}~\bibnamefont {Wang}}, \bibinfo {author} {\bibfnamefont
			{R.~A.}\ \bibnamefont {Ewings}}, \bibinfo {author} {\bibfnamefont
			{T.}~\bibnamefont {Guidi}}, \bibinfo {author} {\bibfnamefont
			{H.}~\bibnamefont {Park}}, \bibinfo {author} {\bibfnamefont {K.}~\bibnamefont
			{Haule}}, \bibinfo {author} {\bibfnamefont {G.}~\bibnamefont {Kotliar}},
		\bibinfo {author} {\bibfnamefont {S.~M.}\ \bibnamefont {Hayden}},\ and\
		\bibinfo {author} {\bibfnamefont {P.}~\bibnamefont {Dai}},\ }\bibfield
	{title} {\bibinfo {title} {Nature of magnetic excitations in superconducting
			{BaFe$_{1.9}$Ni$_{0.1}$As$_2$}},\ }\href {https://doi.org/10.1038/nphys2268}
	{\bibfield  {journal} {\bibinfo  {journal} {Nat. Phys.}\ }\textbf {\bibinfo
			{volume} {8}},\ \bibinfo {pages} {376} (\bibinfo {year}
		{2012}{\natexlab{b}})}\BibitemShut {NoStop}%
	\bibitem [{\citenamefont {Luo}\ \emph {et~al.}(2013)\citenamefont {Luo},
		\citenamefont {Lu}, \citenamefont {Zhang}, \citenamefont {Wang},
		\citenamefont {Goremychkin}, \citenamefont {Adroja}, \citenamefont
		{Danilkin}, \citenamefont {Deng}, \citenamefont {Yamani},\ and\ \citenamefont
		{Dai}}]{LuoBaNi}%
	\BibitemOpen
	\bibfield  {author} {\bibinfo {author} {\bibfnamefont {H.}~\bibnamefont
			{Luo}}, \bibinfo {author} {\bibfnamefont {X.}~\bibnamefont {Lu}}, \bibinfo
		{author} {\bibfnamefont {R.}~\bibnamefont {Zhang}}, \bibinfo {author}
		{\bibfnamefont {M.}~\bibnamefont {Wang}}, \bibinfo {author} {\bibfnamefont
			{E.~A.}\ \bibnamefont {Goremychkin}}, \bibinfo {author} {\bibfnamefont
			{D.~T.}\ \bibnamefont {Adroja}}, \bibinfo {author} {\bibfnamefont
			{S.}~\bibnamefont {Danilkin}}, \bibinfo {author} {\bibfnamefont
			{G.}~\bibnamefont {Deng}}, \bibinfo {author} {\bibfnamefont {Z.}~\bibnamefont
			{Yamani}},\ and\ \bibinfo {author} {\bibfnamefont {P.}~\bibnamefont {Dai}},\
	}\bibfield  {title} {\bibinfo {title} {Electron doping evolution of the
			magnetic excitations in {BaFe$_{2-\textit{x}}$Ni$_\textit{x}$As$_2$}},\
	}\href {https://doi.org/10.1103/PhysRevB.88.144516} {\bibfield  {journal}
		{\bibinfo  {journal} {Phys. Rev. B}\ }\textbf {\bibinfo {volume} {88}},\
		\bibinfo {pages} {144516} (\bibinfo {year} {2013})}\BibitemShut {NoStop}%
	\bibitem [{\citenamefont {Wang}\ \emph {et~al.}(2013)\citenamefont {Wang},
		\citenamefont {Zhang}, \citenamefont {Lu}, \citenamefont {Tan}, \citenamefont
		{Luo}, \citenamefont {Song}, \citenamefont {Wang}, \citenamefont {Zhang},
		\citenamefont {Goremychkin}, \citenamefont {Perring}, \citenamefont {Maier},
		\citenamefont {Yin}, \citenamefont {Haule}, \citenamefont {Kotliar},\ and\
		\citenamefont {Dai}}]{MengWang}%
	\BibitemOpen
	\bibfield  {author} {\bibinfo {author} {\bibfnamefont {M.}~\bibnamefont
			{Wang}}, \bibinfo {author} {\bibfnamefont {C.}~\bibnamefont {Zhang}},
		\bibinfo {author} {\bibfnamefont {X.}~\bibnamefont {Lu}}, \bibinfo {author}
		{\bibfnamefont {G.}~\bibnamefont {Tan}}, \bibinfo {author} {\bibfnamefont
			{H.}~\bibnamefont {Luo}}, \bibinfo {author} {\bibfnamefont {Y.}~\bibnamefont
			{Song}}, \bibinfo {author} {\bibfnamefont {M.}~\bibnamefont {Wang}}, \bibinfo
		{author} {\bibfnamefont {X.}~\bibnamefont {Zhang}}, \bibinfo {author}
		{\bibfnamefont {E.~A.}\ \bibnamefont {Goremychkin}}, \bibinfo {author}
		{\bibfnamefont {T.~G.}\ \bibnamefont {Perring}}, \bibinfo {author}
		{\bibfnamefont {T.~A.}\ \bibnamefont {Maier}}, \bibinfo {author}
		{\bibfnamefont {Z.}~\bibnamefont {Yin}}, \bibinfo {author} {\bibfnamefont
			{K.}~\bibnamefont {Haule}}, \bibinfo {author} {\bibfnamefont
			{G.}~\bibnamefont {Kotliar}},\ and\ \bibinfo {author} {\bibfnamefont
			{P.}~\bibnamefont {Dai}},\ }\bibfield  {title} {\bibinfo {title} {Doping
			dependence of spin excitations and its correlations with high-temperature
			superconductivity in iron pnictides},\ }\href
	{https://doi.org/10.1038/ncomms3874} {\bibfield  {journal} {\bibinfo
			{journal} {Nat. Commun.}\ }\textbf {\bibinfo {volume} {4}},\ \bibinfo {pages}
		{2874} (\bibinfo {year} {2013})}\BibitemShut {NoStop}%
	\bibitem [{\citenamefont {Xie}\ \emph {et~al.}(2022)\citenamefont {Xie},
		\citenamefont {Liu}, \citenamefont {Kajimoto}, \citenamefont {Ikeuchi},
		\citenamefont {Li},\ and\ \citenamefont {Luo}}]{112_Luo}%
	\BibitemOpen
	\bibfield  {author} {\bibinfo {author} {\bibfnamefont {T.}~\bibnamefont
			{Xie}}, \bibinfo {author} {\bibfnamefont {C.}~\bibnamefont {Liu}}, \bibinfo
		{author} {\bibfnamefont {R.}~\bibnamefont {Kajimoto}}, \bibinfo {author}
		{\bibfnamefont {K.}~\bibnamefont {Ikeuchi}}, \bibinfo {author} {\bibfnamefont
			{S.}~\bibnamefont {Li}},\ and\ \bibinfo {author} {\bibfnamefont
			{H.}~\bibnamefont {Luo}},\ }\bibfield  {title} {\bibinfo {title} {Spin
			fluctuations in the 112-type iron-based superconductor
			{Ca$_{0.82}$La$_{0.18}$Fe$_{0.96}$Ni$_{0.04}$As$_2$}},\ }\href
	{https://doi.org/10.1088/1361-648X/ac9441} {\bibfield  {journal} {\bibinfo
			{journal} {J. Phys.: Condens. Matter}\ }\textbf {\bibinfo {volume} {34}},\
		\bibinfo {pages} {474001} (\bibinfo {year} {2022})}\BibitemShut {NoStop}%
	\bibitem [{\citenamefont {Lorenzana}\ \emph {et~al.}(2005)\citenamefont
		{Lorenzana}, \citenamefont {Seibold},\ and\ \citenamefont
		{Coldea}}]{MissingSpec}%
	\BibitemOpen
	\bibfield  {author} {\bibinfo {author} {\bibfnamefont {J.}~\bibnamefont
			{Lorenzana}}, \bibinfo {author} {\bibfnamefont {G.}~\bibnamefont {Seibold}},\
		and\ \bibinfo {author} {\bibfnamefont {R.}~\bibnamefont {Coldea}},\
	}\bibfield  {title} {\bibinfo {title} {Sum rules and missing spectral weight
			in magnetic neutron scattering in the cuprates},\ }\href
	{https://doi.org/10.1103/PhysRevB.72.224511} {\bibfield  {journal} {\bibinfo
			{journal} {Phys. Rev. B}\ }\textbf {\bibinfo {volume} {72}},\ \bibinfo
		{pages} {224511} (\bibinfo {year} {2005})}\BibitemShut {NoStop}%
	\bibitem [{\citenamefont {Lee}\ \emph {et~al.}(2006)\citenamefont {Lee},
		\citenamefont {Nagaosa},\ and\ \citenamefont {Wen}}]{CuprateRMP}%
	\BibitemOpen
	\bibfield  {author} {\bibinfo {author} {\bibfnamefont {P.~A.}\ \bibnamefont
			{Lee}}, \bibinfo {author} {\bibfnamefont {N.}~\bibnamefont {Nagaosa}},\ and\
		\bibinfo {author} {\bibfnamefont {X.~G.}\ \bibnamefont {Wen}},\ }\bibfield
	{title} {\bibinfo {title} {Doping a {Mott} insulator: {Physics} of
			high-temperature superconductivity},\ }\href
	{https://doi.org/10.1103/RevModPhys.78.17} {\bibfield  {journal} {\bibinfo
			{journal} {Rev. Mod. Phys.}\ }\textbf {\bibinfo {volume} {78}},\ \bibinfo
		{pages} {17} (\bibinfo {year} {2006})}\BibitemShut {NoStop}%
	\bibitem [{\citenamefont {Hayden}\ \emph {et~al.}(2004)\citenamefont {Hayden},
		\citenamefont {Mook}, \citenamefont {Dai}, \citenamefont {Perring},\ and\
		\citenamefont {Dogan}}]{Hourglass1}%
	\BibitemOpen
	\bibfield  {author} {\bibinfo {author} {\bibfnamefont {S.~M.}\ \bibnamefont
			{Hayden}}, \bibinfo {author} {\bibfnamefont {H.~A.}\ \bibnamefont {Mook}},
		\bibinfo {author} {\bibfnamefont {P.~C.}\ \bibnamefont {Dai}}, \bibinfo
		{author} {\bibfnamefont {T.~G.}\ \bibnamefont {Perring}},\ and\ \bibinfo
		{author} {\bibfnamefont {F.}~\bibnamefont {Dogan}},\ }\bibfield  {title}
	{\bibinfo {title} {The structure of the high-energy spin excitations in a
			high-transition-temperature superconductor},\ }\href
	{https://doi.org/10.1038/nature02576} {\bibfield  {journal} {\bibinfo
			{journal} {Nature}\ }\textbf {\bibinfo {volume} {429}},\ \bibinfo {pages}
		{531} (\bibinfo {year} {2004})}\BibitemShut {NoStop}%
	\bibitem [{\citenamefont {Tranquada}\ \emph {et~al.}(2004)\citenamefont
		{Tranquada}, \citenamefont {Woo}, \citenamefont {Perring}, \citenamefont
		{Goka}, \citenamefont {Gu}, \citenamefont {Xu}, \citenamefont {Fujita},\ and\
		\citenamefont {Yamada}}]{Hourglass2}%
	\BibitemOpen
	\bibfield  {author} {\bibinfo {author} {\bibfnamefont {J.~M.}\ \bibnamefont
			{Tranquada}}, \bibinfo {author} {\bibfnamefont {H.}~\bibnamefont {Woo}},
		\bibinfo {author} {\bibfnamefont {T.~G.}\ \bibnamefont {Perring}}, \bibinfo
		{author} {\bibfnamefont {H.}~\bibnamefont {Goka}}, \bibinfo {author}
		{\bibfnamefont {G.~D.}\ \bibnamefont {Gu}}, \bibinfo {author} {\bibfnamefont
			{G.}~\bibnamefont {Xu}}, \bibinfo {author} {\bibfnamefont {M.}~\bibnamefont
			{Fujita}},\ and\ \bibinfo {author} {\bibfnamefont {K.}~\bibnamefont
			{Yamada}},\ }\bibfield  {title} {\bibinfo {title} {Quantum magnetic
			excitations from stripes in copper oxide superconductors},\ }\href
	{https://doi.org/10.1038/nature02574} {\bibfield  {journal} {\bibinfo
			{journal} {Nature}\ }\textbf {\bibinfo {volume} {429}},\ \bibinfo {pages}
		{534} (\bibinfo {year} {2004})}\BibitemShut {NoStop}%
	\bibitem [{\citenamefont {Vignolle}\ \emph {et~al.}(2007)\citenamefont
		{Vignolle}, \citenamefont {Hayden}, \citenamefont {McMorrow}, \citenamefont
		{Rønnow}, \citenamefont {Lake}, \citenamefont {Frost},\ and\ \citenamefont
		{Perring}}]{Hourglass3}%
	\BibitemOpen
	\bibfield  {author} {\bibinfo {author} {\bibfnamefont {B.}~\bibnamefont
			{Vignolle}}, \bibinfo {author} {\bibfnamefont {S.~M.}\ \bibnamefont
			{Hayden}}, \bibinfo {author} {\bibfnamefont {D.~F.}\ \bibnamefont
			{McMorrow}}, \bibinfo {author} {\bibfnamefont {H.~M.}\ \bibnamefont
			{Rønnow}}, \bibinfo {author} {\bibfnamefont {B.}~\bibnamefont {Lake}},
		\bibinfo {author} {\bibfnamefont {C.~D.}\ \bibnamefont {Frost}},\ and\
		\bibinfo {author} {\bibfnamefont {T.~G.}\ \bibnamefont {Perring}},\
	}\bibfield  {title} {\bibinfo {title} {Two energy scales in the spin
			excitations of the high-temperature superconductor
			{La$_{2-x}$Sr$_x$CuO$_4$}},\ }\href {https://doi.org/10.1038/nphys546}
	{\bibfield  {journal} {\bibinfo  {journal} {Nat. Phys.}\ }\textbf {\bibinfo
			{volume} {3}},\ \bibinfo {pages} {163} (\bibinfo {year} {2007})}\BibitemShut
	{NoStop}%
	\bibitem [{\citenamefont {Bulut}\ \emph {et~al.}(1990)\citenamefont {Bulut},
		\citenamefont {Hone}, \citenamefont {Scalapino},\ and\ \citenamefont
		{Bickers}}]{Bulut}%
	\BibitemOpen
	\bibfield  {author} {\bibinfo {author} {\bibfnamefont {N.}~\bibnamefont
			{Bulut}}, \bibinfo {author} {\bibfnamefont {D.}~\bibnamefont {Hone}},
		\bibinfo {author} {\bibfnamefont {D.~J.}\ \bibnamefont {Scalapino}},\ and\
		\bibinfo {author} {\bibfnamefont {N.~E.}\ \bibnamefont {Bickers}},\
	}\bibfield  {title} {\bibinfo {title} {Random-phase-approximation analysis of
			{NMR} and neutron-scattering experiments on layered cuprates},\ }\href
	{https://doi.org/10.1103/PhysRevLett.64.2723} {\bibfield  {journal} {\bibinfo
			{journal} {Phys. Rev. Lett.}\ }\textbf {\bibinfo {volume} {64}},\ \bibinfo
		{pages} {2723} (\bibinfo {year} {1990})}\BibitemShut {NoStop}%
	\bibitem [{\citenamefont {Si}\ \emph {et~al.}(1993)\citenamefont {Si},
		\citenamefont {Zha}, \citenamefont {Levin},\ and\ \citenamefont {Lu}}]{QMSi}%
	\BibitemOpen
	\bibfield  {author} {\bibinfo {author} {\bibfnamefont {Q.~M.}\ \bibnamefont
			{Si}}, \bibinfo {author} {\bibfnamefont {Y.~Y.}\ \bibnamefont {Zha}},
		\bibinfo {author} {\bibfnamefont {K.}~\bibnamefont {Levin}},\ and\ \bibinfo
		{author} {\bibfnamefont {J.~P.}\ \bibnamefont {Lu}},\ }\bibfield  {title}
	{\bibinfo {title} {Comparison of spin dynamics in
			{YBa$_2$Cu$_3$O$_{7-\delta}$} and {La$_{2-x}$Sr$_x$CuO$_4$}: {Effects} of
			{Fermi}-surface geometry},\ }\href {https://doi.org/10.1103/PhysRevB.47.9055}
	{\bibfield  {journal} {\bibinfo  {journal} {Phys. Rev. B}\ }\textbf {\bibinfo
			{volume} {47}},\ \bibinfo {pages} {9055} (\bibinfo {year}
		{1993})}\BibitemShut {NoStop}%
	\bibitem [{\citenamefont {Littlewood}\ \emph {et~al.}(1993)\citenamefont
		{Littlewood}, \citenamefont {Zaanen}, \citenamefont {Aeppli},\ and\
		\citenamefont {Monien}}]{Littlewood}%
	\BibitemOpen
	\bibfield  {author} {\bibinfo {author} {\bibfnamefont {P.~B.}\ \bibnamefont
			{Littlewood}}, \bibinfo {author} {\bibfnamefont {J.}~\bibnamefont {Zaanen}},
		\bibinfo {author} {\bibfnamefont {G.}~\bibnamefont {Aeppli}},\ and\ \bibinfo
		{author} {\bibfnamefont {H.}~\bibnamefont {Monien}},\ }\bibfield  {title}
	{\bibinfo {title} {Spin fluctuations in a two-dimensional marginal {Fermi}
			liquid},\ }\href {https://doi.org/10.1103/PhysRevB.48.487} {\bibfield
		{journal} {\bibinfo  {journal} {Phys. Rev. B}\ }\textbf {\bibinfo {volume}
			{48}},\ \bibinfo {pages} {487} (\bibinfo {year} {1993})}\BibitemShut
	{NoStop}%
	\bibitem [{\citenamefont {Norman}(2000)}]{Norman}%
	\BibitemOpen
	\bibfield  {author} {\bibinfo {author} {\bibfnamefont {M.~R.}\ \bibnamefont
			{Norman}},\ }\bibfield  {title} {\bibinfo {title} {Relation of neutron
			incommensurability to electronic structure in high-temperature
			superconductors},\ }\href {https://doi.org/10.1103/PhysRevB.61.14751}
	{\bibfield  {journal} {\bibinfo  {journal} {Phys. Rev. B}\ }\textbf {\bibinfo
			{volume} {61}},\ \bibinfo {pages} {14751} (\bibinfo {year}
		{2000})}\BibitemShut {NoStop}%
	\bibitem [{\citenamefont {Hussey}\ \emph {et~al.}(2003)\citenamefont {Hussey},
		\citenamefont {Abdel-Jawad}, \citenamefont {Carrington}, \citenamefont
		{Mackenzie},\ and\ \citenamefont {Balicas}}]{Hussey}%
	\BibitemOpen
	\bibfield  {author} {\bibinfo {author} {\bibfnamefont {N.~E.}\ \bibnamefont
			{Hussey}}, \bibinfo {author} {\bibfnamefont {M.}~\bibnamefont {Abdel-Jawad}},
		\bibinfo {author} {\bibfnamefont {A.}~\bibnamefont {Carrington}}, \bibinfo
		{author} {\bibfnamefont {A.~P.}\ \bibnamefont {Mackenzie}},\ and\ \bibinfo
		{author} {\bibfnamefont {L.}~\bibnamefont {Balicas}},\ }\bibfield  {title}
	{\bibinfo {title} {A coherent three-dimensional {Fermi} surface in a
			high-transition-temperature superconductor},\ }\href
	{https://doi.org/10.1038/nature01981} {\bibfield  {journal} {\bibinfo
			{journal} {Nature}\ }\textbf {\bibinfo {volume} {425}},\ \bibinfo {pages}
		{814} (\bibinfo {year} {2003})}\BibitemShut {NoStop}%
	\bibitem [{\citenamefont {Tranquada}\ \emph {et~al.}(1995)\citenamefont
		{Tranquada}, \citenamefont {Sternlieb}, \citenamefont {Axe}, \citenamefont
		{Nakamura},\ and\ \citenamefont {Uchida}}]{stripe1}%
	\BibitemOpen
	\bibfield  {author} {\bibinfo {author} {\bibfnamefont {J.~M.}\ \bibnamefont
			{Tranquada}}, \bibinfo {author} {\bibfnamefont {B.~J.}\ \bibnamefont
			{Sternlieb}}, \bibinfo {author} {\bibfnamefont {J.~D.}\ \bibnamefont {Axe}},
		\bibinfo {author} {\bibfnamefont {Y.}~\bibnamefont {Nakamura}},\ and\
		\bibinfo {author} {\bibfnamefont {S.}~\bibnamefont {Uchida}},\ }\bibfield
	{title} {\bibinfo {title} {Evidence for stripe correlations of spins and
			holes in copper oxide superconductors},\ }\href
	{https://doi.org/10.1038/375561a0} {\bibfield  {journal} {\bibinfo  {journal}
			{Nature}\ }\textbf {\bibinfo {volume} {375}},\ \bibinfo {pages} {561}
		(\bibinfo {year} {1995})}\BibitemShut {NoStop}%
	\bibitem [{\citenamefont {Kivelson}\ \emph {et~al.}(2003)\citenamefont
		{Kivelson}, \citenamefont {Bindloss}, \citenamefont {Fradkin}, \citenamefont
		{Oganesyan}, \citenamefont {Tranquada}, \citenamefont {Kapitulnik},\ and\
		\citenamefont {Howald}}]{stripe2}%
	\BibitemOpen
	\bibfield  {author} {\bibinfo {author} {\bibfnamefont {S.~A.}\ \bibnamefont
			{Kivelson}}, \bibinfo {author} {\bibfnamefont {I.~P.}\ \bibnamefont
			{Bindloss}}, \bibinfo {author} {\bibfnamefont {E.}~\bibnamefont {Fradkin}},
		\bibinfo {author} {\bibfnamefont {V.}~\bibnamefont {Oganesyan}}, \bibinfo
		{author} {\bibfnamefont {J.~M.}\ \bibnamefont {Tranquada}}, \bibinfo {author}
		{\bibfnamefont {A.}~\bibnamefont {Kapitulnik}},\ and\ \bibinfo {author}
		{\bibfnamefont {C.}~\bibnamefont {Howald}},\ }\bibfield  {title} {\bibinfo
		{title} {How to detect fluctuating stripes in the high-temperature
			superconductors},\ }\href {https://doi.org/10.1103/RevModPhys.75.1201}
	{\bibfield  {journal} {\bibinfo  {journal} {Rev. Mod. Phys.}\ }\textbf
		{\bibinfo {volume} {75}},\ \bibinfo {pages} {1201} (\bibinfo {year}
		{2003})}\BibitemShut {NoStop}%
	\bibitem [{\citenamefont {Wilson}\ \emph {et~al.}(2006)\citenamefont {Wilson},
		\citenamefont {Li}, \citenamefont {Woo}, \citenamefont {Dai}, \citenamefont
		{Mook}, \citenamefont {Frost}, \citenamefont {Komiya},\ and\ \citenamefont
		{Ando}}]{Wilson_PLCCO}%
	\BibitemOpen
	\bibfield  {author} {\bibinfo {author} {\bibfnamefont {S.~D.}\ \bibnamefont
			{Wilson}}, \bibinfo {author} {\bibfnamefont {S.}~\bibnamefont {Li}}, \bibinfo
		{author} {\bibfnamefont {H.}~\bibnamefont {Woo}}, \bibinfo {author}
		{\bibfnamefont {P.}~\bibnamefont {Dai}}, \bibinfo {author} {\bibfnamefont
			{H.~A.}\ \bibnamefont {Mook}}, \bibinfo {author} {\bibfnamefont {C.~D.}\
			\bibnamefont {Frost}}, \bibinfo {author} {\bibfnamefont {S.}~\bibnamefont
			{Komiya}},\ and\ \bibinfo {author} {\bibfnamefont {Y.}~\bibnamefont {Ando}},\
	}\bibfield  {title} {\bibinfo {title} {High-energy spin excitations in the
			electron-doped superconductor {Pr$_{0.88}$LaCe$_{0.12}$CuO$_{4-\delta}$} with
			\textit{{T}}$_c$ = 21 {K}},\ }\href
	{https://doi.org/10.1103/PhysRevLett.96.157001} {\bibfield  {journal}
		{\bibinfo  {journal} {Phys. Rev. Lett.}\ }\textbf {\bibinfo {volume} {96}},\
		\bibinfo {pages} {157001} (\bibinfo {year} {2006})}\BibitemShut {NoStop}%
	\bibitem [{\citenamefont {Zhao}\ \emph {et~al.}(2011)\citenamefont {Zhao},
		\citenamefont {Niestemski}, \citenamefont {Kunwar}, \citenamefont {Li},
		\citenamefont {Steffens}, \citenamefont {Hiess}, \citenamefont {Kang},
		\citenamefont {Wilson}, \citenamefont {Wang}, \citenamefont {Dai},\ and\
		\citenamefont {Madhavan}}]{Jun_PLCCO}%
	\BibitemOpen
	\bibfield  {author} {\bibinfo {author} {\bibfnamefont {J.}~\bibnamefont
			{Zhao}}, \bibinfo {author} {\bibfnamefont {F.~C.}\ \bibnamefont
			{Niestemski}}, \bibinfo {author} {\bibfnamefont {S.}~\bibnamefont {Kunwar}},
		\bibinfo {author} {\bibfnamefont {S.}~\bibnamefont {Li}}, \bibinfo {author}
		{\bibfnamefont {P.}~\bibnamefont {Steffens}}, \bibinfo {author}
		{\bibfnamefont {A.}~\bibnamefont {Hiess}}, \bibinfo {author} {\bibfnamefont
			{H.~J.}\ \bibnamefont {Kang}}, \bibinfo {author} {\bibfnamefont {S.~D.}\
			\bibnamefont {Wilson}}, \bibinfo {author} {\bibfnamefont {Z.}~\bibnamefont
			{Wang}}, \bibinfo {author} {\bibfnamefont {P.}~\bibnamefont {Dai}},\ and\
		\bibinfo {author} {\bibfnamefont {V.}~\bibnamefont {Madhavan}},\ }\bibfield
	{title} {\bibinfo {title} {Electron-spin excitation coupling in an
			electron-doped copper oxide superconductor},\ }\href
	{https://doi.org/10.1038/NPHYS2006} {\bibfield  {journal} {\bibinfo
			{journal} {Nat. Phys.}\ }\textbf {\bibinfo {volume} {7}},\ \bibinfo {pages}
		{719} (\bibinfo {year} {2011})}\BibitemShut {NoStop}%
	\bibitem [{\citenamefont {Armitage}\ \emph {et~al.}(2010)\citenamefont
		{Armitage}, \citenamefont {Fournier},\ and\ \citenamefont
		{Greene}}]{ntypeCu}%
	\BibitemOpen
	\bibfield  {author} {\bibinfo {author} {\bibfnamefont {N.~P.}\ \bibnamefont
			{Armitage}}, \bibinfo {author} {\bibfnamefont {P.}~\bibnamefont {Fournier}},\
		and\ \bibinfo {author} {\bibfnamefont {R.~L.}\ \bibnamefont {Greene}},\
	}\bibfield  {title} {\bibinfo {title} {Progress and perspectives on
			electron-doped cuprates},\ }\href
	{https://doi.org/10.1103/RevModPhys.82.2421} {\bibfield  {journal} {\bibinfo
			{journal} {Rev. Mod. Phys.}\ }\textbf {\bibinfo {volume} {82}},\ \bibinfo
		{pages} {2421} (\bibinfo {year} {2010})}\BibitemShut {NoStop}%
	\bibitem [{\citenamefont {Eschrig}(2006)}]{Eschrig}%
	\BibitemOpen
	\bibfield  {author} {\bibinfo {author} {\bibfnamefont {M.}~\bibnamefont
			{Eschrig}},\ }\bibfield  {title} {\bibinfo {title} {The effect of collective
			spin-1 excitations on electronic spectra in high-\textit{{T}}$_c$
			superconductors},\ }\href {https://doi.org/10.1080/00018730600645636}
	{\bibfield  {journal} {\bibinfo  {journal} {Adv. Phys.}\ }\textbf {\bibinfo
			{volume} {55}},\ \bibinfo {pages} {47} (\bibinfo {year} {2006})}\BibitemShut
	{NoStop}%
	\bibitem [{\citenamefont {Wang}\ \emph {et~al.}(2011)\citenamefont {Wang},
		\citenamefont {Yang}, \citenamefont {Gao}, \citenamefont {Lu}, \citenamefont
		{Xiang},\ and\ \citenamefont {Lee}}]{dwave_FeSe1}%
	\BibitemOpen
	\bibfield  {author} {\bibinfo {author} {\bibfnamefont {F.}~\bibnamefont
			{Wang}}, \bibinfo {author} {\bibfnamefont {F.}~\bibnamefont {Yang}}, \bibinfo
		{author} {\bibfnamefont {M.}~\bibnamefont {Gao}}, \bibinfo {author}
		{\bibfnamefont {Z.~Y.}\ \bibnamefont {Lu}}, \bibinfo {author} {\bibfnamefont
			{T.}~\bibnamefont {Xiang}},\ and\ \bibinfo {author} {\bibfnamefont {D.~H.}\
			\bibnamefont {Lee}},\ }\bibfield  {title} {\bibinfo {title} {The electron
			pairing of {K$_\textit{x}$Fe$_{2-\textit{y}}$Se$_2$}},\ }\href
	{https://doi.org/10.1209/0295-5075/93/57003} {\bibfield  {journal} {\bibinfo
			{journal} {Europhysics Letters}\ }\textbf {\bibinfo {volume} {93}},\ \bibinfo
		{pages} {57003} (\bibinfo {year} {2011})}\BibitemShut {NoStop}%
	\bibitem [{\citenamefont {Maier}\ \emph {et~al.}(2011)\citenamefont {Maier},
		\citenamefont {Graser}, \citenamefont {Hirschfeld},\ and\ \citenamefont
		{Scalapino}}]{dwave_FeSe2}%
	\BibitemOpen
	\bibfield  {author} {\bibinfo {author} {\bibfnamefont {T.~A.}\ \bibnamefont
			{Maier}}, \bibinfo {author} {\bibfnamefont {S.}~\bibnamefont {Graser}},
		\bibinfo {author} {\bibfnamefont {P.~J.}\ \bibnamefont {Hirschfeld}},\ and\
		\bibinfo {author} {\bibfnamefont {D.~J.}\ \bibnamefont {Scalapino}},\
	}\bibfield  {title} {\bibinfo {title} {\textit{d}-wave pairing from spin
			fluctuations in the {K$_\textit{x}$Fe$_{2-\textit{y}}$Se$_2$}
			superconductors},\ }\href {https://doi.org/10.1103/PhysRevB.83.100515}
	{\bibfield  {journal} {\bibinfo  {journal} {Phys. Rev. B}\ }\textbf {\bibinfo
			{volume} {83}},\ \bibinfo {pages} {100515} (\bibinfo {year}
		{2011})}\BibitemShut {NoStop}%
	\bibitem [{\citenamefont {Agterberg}\ \emph {et~al.}(2017)\citenamefont
		{Agterberg}, \citenamefont {Shishidou}, \citenamefont {O'Halloran},
		\citenamefont {Brydon},\ and\ \citenamefont {Weinert}}]{dwave_FeSe3}%
	\BibitemOpen
	\bibfield  {author} {\bibinfo {author} {\bibfnamefont {D.~F.}\ \bibnamefont
			{Agterberg}}, \bibinfo {author} {\bibfnamefont {T.}~\bibnamefont
			{Shishidou}}, \bibinfo {author} {\bibfnamefont {J.}~\bibnamefont
			{O'Halloran}}, \bibinfo {author} {\bibfnamefont {P.~M.~R.}\ \bibnamefont
			{Brydon}},\ and\ \bibinfo {author} {\bibfnamefont {M.}~\bibnamefont
			{Weinert}},\ }\bibfield  {title} {\bibinfo {title} {Resilient nodeless
			\textit{d}-wave superconductivity in monolayer {FeSe}},\ }\href
	{https://doi.org/10.1103/PhysRevLett.119.267001} {\bibfield  {journal}
		{\bibinfo  {journal} {Phys. Rev. Lett.}\ }\textbf {\bibinfo {volume} {119}},\
		\bibinfo {pages} {267001} (\bibinfo {year} {2017})}\BibitemShut {NoStop}%
	\bibitem [{\citenamefont {Hirschfeld}\ \emph {et~al.}(2011)\citenamefont
		{Hirschfeld}, \citenamefont {Korshunov},\ and\ \citenamefont
		{Mazin}}]{dwave_FeSe4}%
	\BibitemOpen
	\bibfield  {author} {\bibinfo {author} {\bibfnamefont {P.~J.}\ \bibnamefont
			{Hirschfeld}}, \bibinfo {author} {\bibfnamefont {M.~M.}\ \bibnamefont
			{Korshunov}},\ and\ \bibinfo {author} {\bibfnamefont {I.~I.}\ \bibnamefont
			{Mazin}},\ }\bibfield  {title} {\bibinfo {title} {Gap symmetry and structure
			of {Fe}-based superconductors},\ }\href
	{https://doi.org/10.1088/0034-4885/74/12/124508} {\bibfield  {journal}
		{\bibinfo  {journal} {Rep. Prog. Phys.}\ }\textbf {\bibinfo {volume} {74}},\
		\bibinfo {pages} {124508} (\bibinfo {year} {2011})}\BibitemShut {NoStop}%
	\bibitem [{\citenamefont {Chen}\ \emph {et~al.}(2023)\citenamefont {Chen},
		\citenamefont {Li}, \citenamefont {Lu}, \citenamefont {Liu}, \citenamefont
		{Zhang}, \citenamefont {Li}, \citenamefont {Yin}, \citenamefont {Li},
		\citenamefont {Zhang}, \citenamefont {Dong}, \citenamefont {Yan},\ and\
		\citenamefont {Feng}}]{ChargeOrder}%
	\BibitemOpen
	\bibfield  {author} {\bibinfo {author} {\bibfnamefont {Z.}~\bibnamefont
			{Chen}}, \bibinfo {author} {\bibfnamefont {D.}~\bibnamefont {Li}}, \bibinfo
		{author} {\bibfnamefont {Z.}~\bibnamefont {Lu}}, \bibinfo {author}
		{\bibfnamefont {Y.}~\bibnamefont {Liu}}, \bibinfo {author} {\bibfnamefont
			{J.}~\bibnamefont {Zhang}}, \bibinfo {author} {\bibfnamefont
			{Y.}~\bibnamefont {Li}}, \bibinfo {author} {\bibfnamefont {R.}~\bibnamefont
			{Yin}}, \bibinfo {author} {\bibfnamefont {M.}~\bibnamefont {Li}}, \bibinfo
		{author} {\bibfnamefont {T.}~\bibnamefont {Zhang}}, \bibinfo {author}
		{\bibfnamefont {X.}~\bibnamefont {Dong}}, \bibinfo {author} {\bibfnamefont
			{Y.-J.}\ \bibnamefont {Yan}},\ and\ \bibinfo {author} {\bibfnamefont {D.-L.}\
			\bibnamefont {Feng}},\ }\bibfield  {title} {\bibinfo {title} {Charge order
			driven by multiple-{Q} spin fluctuations in heavily electron-doped iron
			selenide superconductors},\ }\href
	{https://doi.org/10.1038/s41467-023-37792-3} {\bibfield  {journal} {\bibinfo
			{journal} {Nat. Commun.}\ }\textbf {\bibinfo {volume} {14}},\ \bibinfo
		{pages} {2023} (\bibinfo {year} {2023})}\BibitemShut {NoStop}%
	\bibitem [{\citenamefont {Ren}\ \emph {et~al.}(2017)\citenamefont {Ren},
		\citenamefont {Yan}, \citenamefont {Niu}, \citenamefont {Tao}, \citenamefont
		{Hu}, \citenamefont {Peng}, \citenamefont {Xie}, \citenamefont {Zhao},
		\citenamefont {Zhang},\ and\ \citenamefont {Feng}}]{K_dose}%
	\BibitemOpen
	\bibfield  {author} {\bibinfo {author} {\bibfnamefont {M.}~\bibnamefont
			{Ren}}, \bibinfo {author} {\bibfnamefont {Y.}~\bibnamefont {Yan}}, \bibinfo
		{author} {\bibfnamefont {X.}~\bibnamefont {Niu}}, \bibinfo {author}
		{\bibfnamefont {R.}~\bibnamefont {Tao}}, \bibinfo {author} {\bibfnamefont
			{D.}~\bibnamefont {Hu}}, \bibinfo {author} {\bibfnamefont {R.}~\bibnamefont
			{Peng}}, \bibinfo {author} {\bibfnamefont {B.}~\bibnamefont {Xie}}, \bibinfo
		{author} {\bibfnamefont {J.}~\bibnamefont {Zhao}}, \bibinfo {author}
		{\bibfnamefont {T.}~\bibnamefont {Zhang}},\ and\ \bibinfo {author}
		{\bibfnamefont {D.-L.}\ \bibnamefont {Feng}},\ }\bibfield  {title} {\bibinfo
		{title} {Superconductivity across {Lifshitz} transition and anomalous
			insulating state in surface {K–dosed (Li$_{0.8}$Fe$_{0.2}$OH)FeSe}},\
	}\href {https://doi.org/10.1126/sciadv.1603238} {\bibfield  {journal}
		{\bibinfo  {journal} {Sci. Adv.}\ }\textbf {\bibinfo {volume} {3}},\ \bibinfo
		{pages} {e1603238} (\bibinfo {year} {2017})}\BibitemShut {NoStop}%
	\bibitem [{\citenamefont {Kang}\ \emph {et~al.}(2020)\citenamefont {Kang},
		\citenamefont {Shi}, \citenamefont {Li}, \citenamefont {Wang}, \citenamefont
		{Zhang}, \citenamefont {Zhao}, \citenamefont {Li}, \citenamefont {Song},
		\citenamefont {Zheng}, \citenamefont {Nie}, \citenamefont {Wu},\ and\
		\citenamefont {Chen}}]{pseudogap1}%
	\BibitemOpen
	\bibfield  {author} {\bibinfo {author} {\bibfnamefont {B.~L.}\ \bibnamefont
			{Kang}}, \bibinfo {author} {\bibfnamefont {M.~Z.}\ \bibnamefont {Shi}},
		\bibinfo {author} {\bibfnamefont {S.~J.}\ \bibnamefont {Li}}, \bibinfo
		{author} {\bibfnamefont {H.~H.}\ \bibnamefont {Wang}}, \bibinfo {author}
		{\bibfnamefont {Q.}~\bibnamefont {Zhang}}, \bibinfo {author} {\bibfnamefont
			{D.}~\bibnamefont {Zhao}}, \bibinfo {author} {\bibfnamefont {J.}~\bibnamefont
			{Li}}, \bibinfo {author} {\bibfnamefont {D.~W.}\ \bibnamefont {Song}},
		\bibinfo {author} {\bibfnamefont {L.~X.}\ \bibnamefont {Zheng}}, \bibinfo
		{author} {\bibfnamefont {L.~P.}\ \bibnamefont {Nie}}, \bibinfo {author}
		{\bibfnamefont {T.}~\bibnamefont {Wu}},\ and\ \bibinfo {author}
		{\bibfnamefont {X.~H.}\ \bibnamefont {Chen}},\ }\bibfield  {title} {\bibinfo
		{title} {Preformed {Cooper} pairs in layered {FeSe}-based superconductors},\
	}\href {https://doi.org/10.1103/PhysRevLett.125.097003} {\bibfield  {journal}
		{\bibinfo  {journal} {Phys. Rev. Lett.}\ }\textbf {\bibinfo {volume} {125}},\
		\bibinfo {pages} {097003} (\bibinfo {year} {2020})}\BibitemShut {NoStop}%
	\bibitem [{\citenamefont {Xu}\ \emph {et~al.}(2021)\citenamefont {Xu},
		\citenamefont {Rong}, \citenamefont {Wang}, \citenamefont {Wu}, \citenamefont
		{Hu}, \citenamefont {Cai}, \citenamefont {Gao}, \citenamefont {Yan},
		\citenamefont {Li}, \citenamefont {Yin}, \citenamefont {Chen}, \citenamefont
		{Huang}, \citenamefont {Zhu}, \citenamefont {Huang}, \citenamefont {Liu},
		\citenamefont {Xu}, \citenamefont {Zhao},\ and\ \citenamefont
		{Zhou}}]{pseudogap2}%
	\BibitemOpen
	\bibfield  {author} {\bibinfo {author} {\bibfnamefont {Y.}~\bibnamefont
			{Xu}}, \bibinfo {author} {\bibfnamefont {H.}~\bibnamefont {Rong}}, \bibinfo
		{author} {\bibfnamefont {Q.}~\bibnamefont {Wang}}, \bibinfo {author}
		{\bibfnamefont {D.}~\bibnamefont {Wu}}, \bibinfo {author} {\bibfnamefont
			{Y.}~\bibnamefont {Hu}}, \bibinfo {author} {\bibfnamefont {Y.}~\bibnamefont
			{Cai}}, \bibinfo {author} {\bibfnamefont {Q.}~\bibnamefont {Gao}}, \bibinfo
		{author} {\bibfnamefont {H.}~\bibnamefont {Yan}}, \bibinfo {author}
		{\bibfnamefont {C.}~\bibnamefont {Li}}, \bibinfo {author} {\bibfnamefont
			{C.}~\bibnamefont {Yin}}, \bibinfo {author} {\bibfnamefont {H.}~\bibnamefont
			{Chen}}, \bibinfo {author} {\bibfnamefont {J.}~\bibnamefont {Huang}},
		\bibinfo {author} {\bibfnamefont {Z.}~\bibnamefont {Zhu}}, \bibinfo {author}
		{\bibfnamefont {Y.}~\bibnamefont {Huang}}, \bibinfo {author} {\bibfnamefont
			{G.}~\bibnamefont {Liu}}, \bibinfo {author} {\bibfnamefont {Z.}~\bibnamefont
			{Xu}}, \bibinfo {author} {\bibfnamefont {L.}~\bibnamefont {Zhao}},\ and\
		\bibinfo {author} {\bibfnamefont {X.~J.}\ \bibnamefont {Zhou}},\ }\bibfield
	{title} {\bibinfo {title} {Spectroscopic evidence of superconductivity
			pairing at {83 K} in single-layer {FeSe/SrTiO$_3$} films},\ }\href
	{https://doi.org/10.1038/s41467-021-23106-y} {\bibfield  {journal} {\bibinfo
			{journal} {Nat. Commun.}\ }\textbf {\bibinfo {volume} {12}},\ \bibinfo
		{pages} {2840} (\bibinfo {year} {2021})}\BibitemShut {NoStop}%
	\bibitem [{\citenamefont {Faeth}\ \emph {et~al.}(2021)\citenamefont {Faeth},
		\citenamefont {Yang}, \citenamefont {Kawasaki}, \citenamefont {Nelson},
		\citenamefont {Mishra}, \citenamefont {Parzyck}, \citenamefont {Li},
		\citenamefont {Schlom},\ and\ \citenamefont {Shen}}]{pseudogap3}%
	\BibitemOpen
	\bibfield  {author} {\bibinfo {author} {\bibfnamefont {B.~D.}\ \bibnamefont
			{Faeth}}, \bibinfo {author} {\bibfnamefont {S.-L.}\ \bibnamefont {Yang}},
		\bibinfo {author} {\bibfnamefont {J.~K.}\ \bibnamefont {Kawasaki}}, \bibinfo
		{author} {\bibfnamefont {J.~N.}\ \bibnamefont {Nelson}}, \bibinfo {author}
		{\bibfnamefont {P.}~\bibnamefont {Mishra}}, \bibinfo {author} {\bibfnamefont
			{C.~T.}\ \bibnamefont {Parzyck}}, \bibinfo {author} {\bibfnamefont
			{C.}~\bibnamefont {Li}}, \bibinfo {author} {\bibfnamefont {D.~G.}\
			\bibnamefont {Schlom}},\ and\ \bibinfo {author} {\bibfnamefont {K.~M.}\
			\bibnamefont {Shen}},\ }\bibfield  {title} {\bibinfo {title} {Incoherent
			{Cooper} pairing and pseudogap behavior in single-layer {FeSe/SrTiO$_3$}},\
	}\href {https://doi.org/10.1103/PhysRevX.11.021054} {\bibfield  {journal}
		{\bibinfo  {journal} {Phys. Rev. X}\ }\textbf {\bibinfo {volume} {11}},\
		\bibinfo {pages} {021054} (\bibinfo {year} {2021})}\BibitemShut {NoStop}%
\end{thebibliography}
\end{document}